\documentclass[aps,prx,superscriptaddress,singlecolumn,tightenlines,showpacs,floatfix,amsmath,amssymb,pdftex]{revtex4-1}
\usepackage[english]{babel}
\usepackage{mathtools}
\usepackage{amsmath}
\usepackage{amssymb}
\usepackage{braket}
\usepackage[normalem]{ulem}
\usepackage[left=2.5cm,right=2.5cm,top=2.5cm,bottom=2.5cm]{geometry}
\usepackage[pdftex,dvipsnames]{xcolor}
\usepackage{todonotes}
\usepackage{longtable}
\usepackage{theorem}
\definecolor{webgreen}{rgb}{0, 0.5, 0}
\definecolor{webblue}{rgb}{0, 0, 0.5}
\definecolor{webred}{rgb}{0.5, 0, 0}
\usepackage{url}
\usepackage{setspace}
\usepackage{amsmath,amssymb,amscd}
\usepackage{amsfonts}
\usepackage{gensymb}
\usepackage{color}
\usepackage{graphicx}
\usepackage{braket}

\begin{document}
\title{The translational side of topological band insulators}
\author{Robert-Jan Slager}
\affiliation{Max-Planck-Institut f\"ur Physik komplexer Systeme, 01187 Dresden, Germany}

\begin{abstract}
Spin-orbit coupled materials have attracted revived prominent research interest as of late, especially due their direct connection with topological notions. Arguably, a hallmark of this pursuit is formed by the concept of the topological band insulator (TBI). In these incompressible systems band inversions, often driven by strong spin-orbit coupling,  result in a state that is topologically distinct from the usual insulator  as long as time reversal symmetry is maintained. More generally, topological states that arise by virtue of a protecting symmetry have resulted in a flourishing research field on both the experimental as well as theoretical side of the condensed matter agenda.
As a prime signature  topological band insulators can exhibit protected spin filtered edge states, whose time reversal invariant partner is spatially separated on the other edge. While these edge states have resultantly been identified with many exotic physical phenomena, we here wish to provide a perspective on the equally rich, although far less explored, aspects of the bulk.

In particular, it turns out that time reversal invariant $\pi$ fluxes can unambiguously probe the nontrivial nature of TBIs in the bulk. That is, such localized fluxes bind modes that culminate in spin-charge separated excitations. Whereas these modes may be perceived as rather artificial from a experimental perspective, the effects can however be mimicked by ubiquitously present dislocations.  Moreover, the formation criteria can in the latter case directly traced back to the status of the dislocation as being the unique probe of the translational symmetry breaking of the underlying crystal. As a result, the study of dislocations pinpoints an indexing of TBIs beyond the standard tenfold way, which in turn is directly signified by the response to dislocations.

Furthermore, the predicted dislocation physics may also be perceived compelling in its own right. Concretely, the retrieved modes share some of the properties of usual edge states, like the spin-filtered nature, but on the other hand are localized to point defects in 2D TBIs or line defects in 3D TBIs. Consequently, these features culminate in novel mechanisms to be exploited. For example, arrays of dislocations in extended defects, as realizable in ordinary grain boundaries, can be shown to results in self-organized semimetals that have distinct transport properties. Most importantly, however, research efforts on the material side as well as  recent experimental signatures further indicate that these ideas can very well provide for an attractive, timely and experimentally viable research agenda beyond the edge state focussed activities.

\end{abstract}
\maketitle
\section{Introduction}
Topology has been serving as a perpetual theme throughout several branches of science. By virtue of many illustrious mathematical pursuits, the simple idea of classifying geometrical objects according to whether they can be smoothly deformed into each other, as often illustrated by the topological equivalence of a coffee mug and donut,  rapidly developed into many mathematical sub fields that intricately connect geometric, algebraic and analytical notions. 
As somewhat of an empirical truth, such rich mathematical subjects of study often resurface as deeper insights of effective physical models and topology has proven to be anything but an exception. Indeed, relations to topological notions have been identified in many different areas of physics and, with the discovery of the quantum Hall effects in particular \cite{Klitzing1980, Tsui1982}, topology has provided for a prominent research theme within the field of condensed matter. In such quantum Hall systems the physical observable, the quantized Hall conductance  \cite{Thouless1982, Niu1985, Haldane1988}, is directly related to a topological invariant that quantifies the topological characterization, thereby playing a role analogous to that of an order parameter in conventional Ginzburg-Landau symmetry breaking based descriptions \cite{Ginzburg1950, Landau1980}. 

More recently it was uncovered that the notions of symmetry and topology can in fact be flourishingly combined. Concretely, this resulted in the concept of so-called symmetry protected topological (SPT) phases, which entail incompressible 
 states that cannot be adiabatically deformed to a trivial product state as long as the defining protecting symmetry is maintained \cite{Spt1,Spt2,Spt3}. These states thus have a topological identification by virtue of the presence of a symmetry. Arguably, the most instigating and rapidly experimentally verified example of such a SPT phase is the $\mathbb{Z}_{2}$ topological band insulator (TBI) \cite{RevMod1, RevMod2}.  In such systems the presence of time reversal symmetry (TRS)  underlies the possibility of a new topologically distinct insulator in addition to the conventional one. While a TBI may seem featurelessly gapped, it was quickly realized that the topological nature results in many exciting physical consequences. First of all, by the very definition of its topological entity, the topological regime can only be reached from the trivial state upon a closing of the gap. This motivates the reason why topological insulators were readily identified in narrow gapped semiconductors that exhibit strong spin-orbit coupling to act as a mechanism to invert the low lying bands. Nevertheless, a deeper significance of spin-orbit coupling was actually found earlier in the generalization of the quantum Hall effect \cite{Zhang2001}. That is, in this four dimensional (4D) variant dissipationless spin currents take the role of the usual chiral edge states. Similalry, the surface of a topological insulator, being an interface between two topologically distinct phases, features protected edge states; each surface hosts only a single pair of spin-filtered helical movers that  circumvent the Nielsen-Ninomiya theorem \cite{Nielsen1981a, Nielsen1981b} by living on the boundary. This "halving" of the number of degrees of freedom has in turn profound ramifications, such as the possibility of majorana excitations on junctions with superconductors \cite{Beenakker2013, Elliott2015}, and is directly linked to the identification of the TBI bulk as being a condensed matter realization of the $\theta$-vacuum. This therefore makes TBIs very promising candidates of matter to probe novel physical effects \cite{Qi2008}.

The rapid theoretical and experimental progress on $\mathbb{Z}_{2}$ topological band insulators was swiftly put in a more general perspective by the so-called ten fold way. Specifying Hamiltonians in terms of their time reversal, particle hole and chiral symmetries, ten different classes are obtained  \cite{Altland1997}. Using the mathematical
machinery of K-theory \cite{Kitaev2009} or the consideration of the associated topological field theory and the resulting surface states  \cite{schnyder2008, ryu2010} then allows for the determination of the number of distinct topological phases for each of these classes in spatial dimension $d$ . This periodic table therefore mathematically underpins that the $\mathbb{Z}_{2}$ topological band insulator (class AII) should be regarded
as a particular instance of a topological phase that arises by virtue of symmetry. Nonetheless, it is a fact that such free electron systems necessitate the translational symmetry breaking by a lattice to form a band structure in the first place. Hence, it is of fundamental and natural importance to consider the impact of these underlying crystal symmetries, encoded by the defining space group. While the interplay of space group symmetries and topological phases has proven to account for a very rich subject over the past few years \cite{Fu2011, Chen2012t, Slager2013,sg1,sg2,sg3,sg4, Kruthoff2016, Po2017, Bradlyn2017}, we here wish to highlight a somewhat underlit aspect of this pursuit being the connection with the unique topological monodromy probe \cite{Mon1, Mon2} characterizing the translational symmetry breaking of the lattice -the dislocation. Due to their nature it should not come as a surprise that, already early on \cite{Ran2009}, it was  found that dislocations bind special modes that can probe the 3D weak invariants \cite{Fu2007a, Fu2007b, Moore2007}, as these in fact are the product of an interplay between the translational structure and the action of TRS. Interestingly, however, these ideas can be generalized to treat the 2D and 3D cases on the same footing. Moreover, apart from physically illuminating classification schemes that incorporate the space group symmetries, such dislocation modes also harbor rich physical consequences in their own right. These include, amongst others, spin-charge separated particles and the formation of self-organized semi-metals and should therefore  provide for a possible new interesting agenda beyond the usual edge state physics, especially in the light of recent experimental progress \cite{Hamasaki2017}. 

This concise review is organized as follows. After stipulating some preliminaries concerning weak invariants and their relation with translations, Section \ref{sec::2D} elucidates the unique status of $\pi$ fluxes and dislocations as bulk probes of $\mathbb{Z}_{2}$ topological band insulators. These ideas are mainly based on Refs. \cite{Juricic2012} and \cite{Mesaros2012b}. Nonetheless, we attempt to put these results on a firmer footing by placing them in the context of previous works, such as Refs. \cite{Qi2008scs} and \cite{Ran2008scs}. In particular, this intends to elucidate the direct correspondence of $\pi$ flux modes, and thus effectively dislocation modes, with the formulation of the bulk $\mathbb{Z}_{2}$ invariant. Using this motive, the formation criteria of dislocation modes then instigate an indexing of TBIs beyond the tenfold way (Section \ref{sec::Classification}), which was the subject of \cite{Slager2013}. In essence, this analysis mainly revolves around the way time reversal invariant momenta in the Brillouin zone are linked. On a most fundamental level, however, this work's main purpose was to point out that incorporation of lattice symmetries gives rise to additional topological structure {\it within} the tenfold way, which by now has developed into a very rich field. Accordingly, we also briefly comment on how these ideas pertain to active pursuits. This perspective can then in turn be used to enrich the analysis of the previously found  dislocation modes in 3D TBIs, which was reported in \cite{kbt2014} and forms the subject of Section \ref{sec::Modes3D}. In Section \ref{sec::GBmetals} the outlined viewpoint is then physically exploited.
Concretely, the results of \cite{Slager2016} indicate arrays of dislocation in extended defects can naturally host self-organized semimetals that have distinct transport properties. Finally, before concluding in Section \ref{sec::Conclusions}, we sketch the promising experimental status in Section \ref{sec::Experimental}.

\section{Weak invariants and translations\label{sec::weak}}
To set the stage, let us first consider the weak invariants and their relation to dislocation modes.
In presence of TRS, represented by the operator $\vartheta$, the Hamiltonian $H$ satisfies $\vartheta H(\mathbf{k}) \vartheta^{-1}=H(-\mathbf{k})$. As a result, special Brillouin zone momenta $\Gamma_{i}$ at which $H(\mathbf{k})=H(-\mathbf{k})$ host protected Kramers degeneracies. In a 2D TBI, or quantum spin Hall phase, 
these pairs then essentially switch partners an odd number of times, leading to
an odd number of crossings of the Fermi surface and an associated spectrum that cannot
continuously be deformed into a trivial insulator. This mechanism was elegantly captured by Fu and Kane, who showed that the bulk $\mathbb{Z}_{2}$ invariant $\nu$ may accordingly be formulated in terms of the matrix of overlaps
\begin{equation}\label{eq::w}
w_{mn}=\langle u_{m}(-\mathbf{k})|\vartheta | u_{n}(\mathbf{k})\rangle,
\end{equation}
of the occupied Bloch wavefunctions. Provided that the bands $| u_{n}(\mathbf{k})\rangle$ are chosen continuously, $\nu$ is then expressed as 
\begin{equation}\label{eq::nu}
 (-1)^\nu=\prod_{{\Gamma}_i}\delta_i
\end{equation}
in terms  of the quantities $\delta_i={\text{Pf}[w({\Gamma}_{i})]}/{\sqrt{\text{det}[w({\Gamma}_{i})]}}$ that, by virtue of the antisymmetric nature of $w_{ij}$, take values $\pm1$ at the time reversal invariant (TRI) points $\Gamma_{i}$.

Switching gear to three spatial dimensions, the presence of TRS similarly allows for an insulating band structure that cannot be continuously deformed to the trivial insulating state as long as it is maintained. Although the relation to the 2D case is more subtle \cite{Moore2007}, the defining invariant $\nu_{0}$ physically pertains to whether an even or an odd number of Kramers points is enclosed by
the Fermi surface and can still be defined using Eq. \eqref{eq::nu}, taking into account all 8 TRI momenta. A state with nontrivial $\nu_{0}$ is then referred to as a strong topological insulator. However, three additional  invariants $(\nu_{1},\nu_{2},\nu_{3})$ can now be identified. These remaining three invariants pertain to Miller indices reflecting the two dimensional cuts of
the Brillouin zone in which the Kramers pairs switch partners and thus intimately relate to the action of TRS as well as translational symmetry \cite{Fu2007a, Fu2007b, Moore2007}. That is, TRS works invariantly in 2D cuts, as signified by the very existence of the quantum spin Hall phase, and hence specifying in which cuts of the Brillouin zone the nontrivial behavior occurs is meaningful. As an ultimate consequence, one may in particular consider a stack of quantum spin Hall phases, say in the $\hat{z}$-direction. As an even number of $\delta(\Gamma_{i})$  along the translational invariant direction would necessarily have to have the same sign, this phase is characterized by $\nu_{0}=0$ while planes perpendicular to the $(0,0,1)$-direction exhibit nontrivial behavior. Such phases were dubbed weak topological insulators and have an obvious relation with the two dimensional case.

\begin{figure}[h]
\center
\includegraphics[scale=0.6]{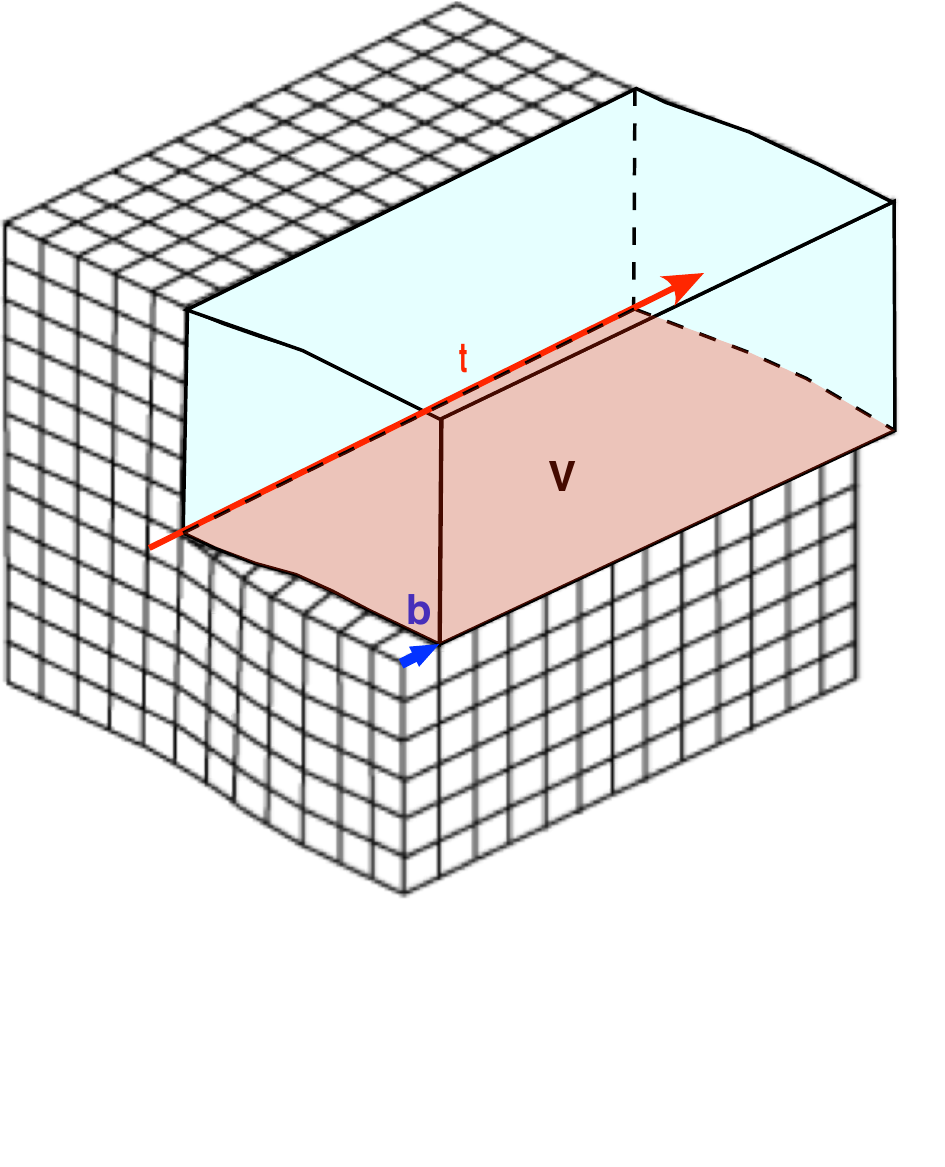}
	\caption{Illustration of the Volterra construction. In this process one singles out a hypothetical plane $V$ and imagines cutting the crystal along this plane. As a next step, the two pieces are figured to be glued back together with a multiple lattice site shift, thereby amounting to a vector known as the Burgers vector ${\bf b}$. The mismatch at the edge of the plane $V$ then results in a line defect, being the red dislocation line that is parametrized by a direction vector ${\bf t}$.}
\label{fig::1}
\end{figure}

Most interestingly, as alluded to, the connection of the weak invariants and translational symmetry can be made physically explicit by studying the response to dislocations as they entail the unique topological probes associated with transitional symmetry. Although we will expose this mechanism in detail below, this was already early on noticed by Ran, Zhang and Vishwanath based on the following insightful argument  \cite{Ran2009}. A dislocation can be thought of as the result of the so-called Volterra process (Figure \ref{fig::1}). In this procedure one singles out a plane and performs a cut along this plane. As a next step, one then glues back the two pieces but with a shift $\mathbf{b}$ that amounts to an integer multiple of lattice sites. As a result, the crystal is left intact except for a defect line at the end of the hypothetical plane, forming the dislocation. Note that the vector representing the shift, the Burgers vector $\mathbf{b}$, can either be perpendicular, creating an edge dislocation, or parallel to the plane, which creates a screw dislocation. Returning to the consideration of the effect of dislocations in TBIs, the shift by $\mathbf{b}$ results in a phase of the hopping amplitude $t\rightarrow te^{i \mathbf{k}\cdot\mathbf{b}}$ across the Volterra plane, where $\mathbf{k}$ is the effective crystal momentum. Now, being a TBI, edges obtained after cutting the whole crystal by extending the Volterra plane host edge states. When glued together these double pairs then gap out with a mass term $m$, as should be the case for the gapped bulk. However, in presence of the dislocation the hoppings acquire the mentioned phase shift across the part comprising the Volterra plane. It is here where the connection to the weak invariants becomes apparent. Namely, as the weak indices signify the wave vector of the Dirac node as $M_{\nu}=1/2(\nu_1G_1+\nu_2G_2+\nu_3G_3)$ in terms of the reciprocal lattice vectors $G_i$, the phase shift $\mathbf{k}\cdot\mathbf{b}$ amounts to $M_{\nu}\cdot\mathbf{b}$. When this phase sums up to 
\begin{equation}\label{eq::3}
M_{\nu}\cdot\mathbf{b}=\pi \text{ mod } 2\pi,
\end{equation}
the gapping mass term $m$ thus changes sign at the defect line. In the end, one therefore obtains an effective Dirac system with a changing mass. Such Dirac Hamiltonians are
well known to lead to low-energy soliton modes that are protected by index theorems. Accordingly, these modes where confirmed by numerical analysis and shown to obey the crucial existential condition on $M_{\nu}$ (Eq. \ref{eq::3}), thereby making the above mentioned connection to translational symmetry explicit \cite{Ran2009, Tteo2010}.  

\section{New lessons from two spatial dimensions\label{sec::2D}}
Although the above 3D results are very appreciable, it is instructive to step back to two spatial dimensions and consider whether these lines of thought can be further extended. After all, there are no weak invariants in 2D. Nonetheless, it turns out that in fact similar perspectives emerge that rest on rather general principles, thereby signaling extensions of the previous arguments. 

\subsection{$\pi$ fluxes modes in 2D topological band insulators}
To make this explicit, let us turn first turn to the, maybe not at first sight, related problem of $\pi$ flux vortices in TBIs. Although the presence of TRS underlies the topological description, it nonetheless constraints fundamentally and especially experimentally the availability of robust probes of this bulk-boundary correspondence or anomalous responses. In particular, for $2+1$ dimensional TBIs the charge Hall response vanishes, and instead a much more involved TRS invariant quantum spin Hall (QSH) effect characterizes the topological phase. It has been shown, however, that a $\pi$-flux vortex, which actually preserves TRS, can play the role of a ${\mathbb Z}_2$ bulk-probe in a QSH insulator through the appearance of topologically protected mid-gap modes \cite{Qi2008scs, Ran2008scs, Juricic2012, Mesaros2012b, Cho2011}.

Concretely, we may consider in this regard the following Dirac Hamiltonian, which besides the ordinary Dirac mass term ($M$) contains a Schr\" odinger kinetic term ($B$),
\begin{equation}\label{eq:cont-ham}
H_{\rm eff}({\bf k})=i\gamma_0\gamma_i k_i+(M-B{\bf k}^2)\gamma_0.
\end{equation}
In the above, the four-dimensional $\gamma$-matrices satisfy the canonical anticommutation relations $\{\gamma_\alpha,\gamma_\beta\}=2\delta_{\mu\nu}$ and are given by $\gamma_0=\tau_3\otimes\sigma_0$, $\gamma_1=\tau_2\otimes\sigma_3$, $\gamma_2=-\tau_1\otimes\sigma_0$. The  Pauli matrices $\{\sigma_0,\sigma_\mu\}$ then act in spin space, with $\tau_0,\sigma_0$ representing the $2\times 2$ identity matrices. 
We furthermore stipulate that the above Hamiltonian \eqref{eq:cont-ham} entails the continuum, i.e. large wavelength limit ($|{\bf k}|\ll 1$), of the well known BHZ model  \cite{Bernevig2006, Koning2007}, 
initially constructed to describe the experimentally verified HgTe quantum well QSH insulator.

To evaluate the effect of a $\pi$-flux,  it is then sufficient to focus on a single spin projection, since the two spin projections are related by the time-reversal operator $T=-i\sigma_2K$, with $K$ being the complex conjugation.
In the lattice counterpart such a $\pi$-flux can readily be incorporated using a Peierls substitution, which amounts to the minimal coupling ${\bf k}\rightarrow{\bf k}+{\bf A}$ in its continuum
version. As result, the Hamiltonian (\ref{eq:cont-ham}) for the spin up component assumes the form
\begin{equation}\label{eq:ham-A}
H_{\rm eff}({\bf k},{\bf A})=\tau_i (k_i+A_i)+[M-B({\bf k}+{\bf A})^2]\tau_3,
\end{equation}
where the vector potential
\begin{equation}\label{eq:A}
{\bf A}=\frac{-y{\bf e}_x+x{\bf e}_y}{2r^2}
\end{equation}
represents the magnetic vortex carrying flux $\Phi=\pi$. 

Let us now show in the remainder of this subsection that the above Hamiltonian possesses precisely one bulk zero-energy state with spin up.
Expressing the Hamiltonian (\ref{eq:ham-A}) in polar coordinates $(r,\varphi)$,  we obtain
\begin{eqnarray}\label{eq:ham-A-explicit}
H_{\rm eff}&=&-i{ e}^{-i\varphi}\left[\partial_r-\frac{i}{r}{\tilde\partial}_\varphi\right]\tau_ {+}-i{ e}^{i\varphi}\left[\partial_r+\frac{i}{r}{\tilde\partial}_\varphi\right]\tau_{-}\nonumber\\
&+&\left[M+B\left(\partial_r^2+\frac{1}{r}\partial_r+\frac{1}{r^2}{\tilde\partial}_\varphi^2\right)\right]\tau_3,
\end{eqnarray}
where ${\tilde\partial}_\varphi\equiv\partial_\varphi+i/2$, and $\tau_\pm\equiv(\tau_1\pm i\tau_2)/2$. It is generally easy to see that in case of an arbitrary flux 
$\Phi$, the Hamiltonian (\ref{eq:ham-A}) also acquires the form  \eqref{eq:ham-A-explicit}, but with the operator ${\tilde\partial}_\varphi=\partial_\varphi+i\Phi/2\pi$.
In the presence of a vortex carrying a $\pi$ flux, we can exploit the well-defined angular momentum and seek for solutions of the form
\begin{equation}\label{ansatz33}
\bm\psi(r,\varphi)=\sum_{l}c_{l}\bm\psi_l(r,\varphi),
\end{equation}
where
\begin{equation}\label{eq:zero-energy-spinor}
\bm\psi_l(r,\varphi)
=
\begin{pmatrix}
e^{il\varphi} u_{l}(r)\\
e^{i(l+1)\varphi}v_{l+1}(r)
\end{pmatrix},
\qquad l\in\mathbb{Z}.
\end{equation}
As a result, the functions $u,v$ are subjected to the following equations
\begin{eqnarray}
&&\Delta_{l+\frac{1}{2}}u_{l}(r)-i\left(\partial_r+\frac{l+\frac{3}{2}}{r}\right)v_{l+1}(r)=0\label{eq:zero-1}\\
&&i\left(\partial_r-\frac{l+\frac{1}{2}}{r}\right)u_{l}(r)+\Delta_{l+\frac{3}{2}}v_{l+1}(r)=0\label{eq:zero-2}.
\end{eqnarray}
Here the operator $\Delta_l$ is defined as
\begin{equation}\label{eq:Delta-def}
\Delta_l\equiv M+B\left(\partial_r^2+\frac{1}{r}\partial_r-\frac{l^2}{r^2}\right)\equiv M+B{O}_l.
\end{equation}
These equation can, after some algebra, be rewritten as
\begin{equation}\label{eq:zero-mode-u}
\left[M^2+(2MB-1){O}_{l+\frac{1}{2}}+B^2{O}_{l+\frac{1}{2}}^2\right]u_{l}(r)=0.
\end{equation}
This result may also be obtained by noting that if the spinor in Eq.\ (\ref{eq:zero-energy-spinor}) is an eigenstate with the zero eigenvalue of the Hamiltonian 
(\ref{eq:ham-A}), then it is also an eigenstate with the same eigenvalue of the square of this Hamiltonian. Using Eq.\ (\ref{eq:ham-A}), one then readily obtains 
\begin{equation}
H_{\rm eff}({\bf k},{\bf A})^2=B^2 ({\tilde{\bf k}}^2)^2+(1-2MB){\tilde{\bf k}}^2+M^2,
\end{equation}
with ${\tilde{\bf k}}\equiv{\bf k}+{\bf A}$, and the operator ${\tilde{\bf k}}^2$ after acting on the angular part of the upper component of the spinor (\ref{eq:zero-energy-spinor}) yields Eq.\ (\ref{eq:zero-mode-u}). Similarly, it may be shown that the function $v_l(r)$ in the spinor given by Eq.\ (\ref{eq:zero-energy-spinor}) obeys an equation of the same form as (\ref{eq:zero-mode-u}) with $l\rightarrow l+1$. From Eq.\ (\ref{eq:zero-mode-u}) we conclude that the function $u_{l}(r)$ is an eigenfunction of the operator ${O}_{l+1/2}$ with a {\it positive} eigenvalue
\begin{equation}\label{eq:u}
{O}_{l+\frac{1}{2}}u_{l}(r)=\lambda^2 u_{l}(r),
\end{equation}
since the operator ${\tilde{\bf k}}^2$ when acting on a function with the angular momentum $l$ is equal to $-{O}_{l+1/2}$ and the eigenstates of the  operator ${\tilde{\bf k}}^2$
with a {\it negative} eigenvalue are localized.  Eqs.\ (\ref{eq:zero-mode-u}) and (\ref{eq:u}) therefore imply
\begin{equation}\label{eq:lambda-square}
\lambda_\pm=\frac{1\pm\sqrt{1-4MB}}{2B},
\end{equation}
and, accordingly, solutions of the form $u_l(r)\sim I_{l+\frac{1}{2}}(\lambda r)$, where $I_l(x)$ is the modified Bessel function of the first kind. However, from the above solutions the only square-integrable ones are those associated with $l-1$ since $I_l(x)\sim x^{-|l|}$ as $x\rightarrow0$. Furthermore, only the linear combination $I_{1/2}(x)-I_{-1/2}(x)\sim x^{-1/2}e^{-x}$ has the asymptotic behavior at infinity consistent with a finite norm of the state.

As a result, we only find square-integrable zero-energy solutions in presence of
the $\pi$-flux vortex. Indeed, there are no normalizable solutions in the absence of the
vortex, due to the asymptotic behavior of the  modified Bessel functions. However, introducing the $\pi$-flux shifts the angular momentum $l$
of the
solutions \eqref{eq:zero-energy-spinor} by a half, resulting in two square integrable zero-energy solutions
in the zero angular momentum channel, being the localized $l=-1$ solutions. In order to make these solutions more explicit, we should distinguish two regimes of parameters, $0<MB<1/4$ and $MB>1/4$, for which the argument of the square-root is positive and negative, respectively.
For $0<MB<1/4$, since the argument of the square-root in the above equation is positive, we obtain two zero-energy solutions
\begin{equation}
  \label{eq:zero-energy-states1}
\bm\Psi_\pm({\bf r})=\frac{e^{-\lambda_\pm r}}{\sqrt{2\pi\lambda_\pm^{-1} r}}\left(\begin{array}{cc}e^{-i\varphi}\\ i\end{array}\right),
\end{equation}
with $\lambda_\pm>0$ because of the square-integrability. On the other hand, when $MB>1/4$, up to a normalization constant, the solutions have the form
\begin{eqnarray}\label{eq:zero-energy-states2}
\bm\Psi_1({\bf r})&=&\frac{e^{-r\sqrt{\frac{M}{B}}\cos\theta}\cos\left(r\sqrt{\frac{M}{B}}\cos\theta\right)}{\sqrt{r}}\left(\begin{array}{cc}e^{-i\varphi}\\ i\end{array}\right),\nonumber\\
\bm\Psi_2({\bf r})&=&\frac{e^{-r\sqrt{\frac{M}{B}}\cos\theta}\sin\left(r\sqrt{\frac{M}{B}}\cos\theta\right)}{\sqrt{r}}\left(\begin{array}{cc}e^{-i\varphi}\\ i\end{array}\right),
\end{eqnarray}
where
\begin{equation}
\theta=\frac{1}{2}\arctan\frac{\sqrt{|1-4MB|}}{1-2MB}.
\end{equation}
However, since the identity
\begin{equation}
  \label{eq:13}
  \sqrt{x}\cos{\left(\frac{1}{2}\arctan\frac{\sqrt{|1-4x|}}{1-2x}\right)}=\frac{1}{2}
\end{equation}
holds for $1/4<x<4$, the localization length of the zero-modes for $MB>1/4$ is actually just a constant independent of $M,B$, namely Eq.~\eqref{eq:zero-energy-states2} becomes
\begin{eqnarray}
  \label{eq:zero-energy-states2final}
\bm\Psi_1({\bf r})&=&\frac{e^{-\frac{r}{2B}}\cos\left(\frac{r}{2B}\right)}{\sqrt{r}}\left(\begin{array}{cc}e^{-i\varphi}\\ i\end{array}\right),\nonumber\\
\bm\Psi_2({\bf r})&=&\frac{e^{-\frac{r}{2B}}\sin\left(\frac{r}{2B}\right)}{\sqrt{r}}\left(\begin{array}{cc}e^{-i\varphi}\\ i\end{array}\right).
\end{eqnarray}
Therefore, we conclude that the Hamiltonian (\ref{eq:ham-A}) possesses zero-energy modes in the entire range of parameters $M$ and $B$ for which the system is in the topologically non-trivially phase \cite{Bernevig2006, Koning2008, Juricic2012}.
Notice also that in the regime when $0<MB<1/4$, there are two characteristic length scales associated with the midgap modes, $\xi_\pm\sim\lambda_\pm^{-1}$. Of course, after a short-distance regularization is imposed, only a linear combination of the two states survives. The physical interpretation of the two length scales depends on the form of the superposition of the state after the regularization has been imposed, as it may be easily seen from the form of the states \eqref{eq:zero-energy-states1}. In the regime $MB>1/4$, the zero-energy states are characterized by a single length-scale $\xi_{\rm loc}\sim 2B$, which is at the same time the localization length and characterizes the oscillations of the exponentially decaying  state.

Furthermore, although the bulk system does not feature normalizable zero-energy solutions, the edge states are closely related to the zero modes bound to the vortex. That is, a derivation analogous to the standard way of obtaining the edge states \cite{Koning2008} reveals that, by imposing open boundary conditions on the wave-functions at one of the edges of the system, the obtained surface states feature a penetration depth given by exactly the same expression as the localization length for the zero-energy  $\pi$-flux vortexmodes. This signals that the bulk-boundary correspondence may thus be probed by inserting a $\pi$-flux vortex in the quantum spin Hall system.

As a final step, the gauge potential then has to be regularized to fix the final form of the single Kramers pair of zero modes; the zero-energy modes, given by Eqs.\ (\ref{eq:zero-energy-states1}) and (\ref{eq:zero-energy-states2final}),
form  an overcomplete basis, reflecting  the fact that the Hamiltonian (\ref{eq:ham-A}) is not self-adjoint due to the singularity of the vortex vector potential (\ref{eq:A}) at the origin.
The simplest possible regularization is provided by considering the vortex with the flux concentrated in a thin annulus of a radius $R$. Considering the Hamiltonian in the range of parameters $0<MB<1/4$ for the moment, the zero-energy state of the Hamiltonian outside the annulus is then a linear combination of the modes $\bm\Psi_\pm$ \ (\ref{eq:zero-energy-states1}) that results from a matching procedure with solutions inside the annulus where  the vector potential ${\bf A}=0$. Taking $R\rightarrow0$, one then obtains, up to a normalization constant, the zero-energy state 
\begin{equation}\label{eq:zero-energy-annulus}
\bm\Psi({\bf r})=\frac{e^{-\lambda_+r}-e^{-\lambda_-r}}{\sqrt{r}}\left(\begin{array}{cc}e^{-i\varphi}\\ i\end{array}\right).
\end{equation}
Notice that this zero-energy state is regular at the origin which is a consequence of the regularity of the vortex-free problem. Similarly, it can be shown that when $MB>1/4$ the zero-energy mode is given by the spinor $\Psi_2$ in Eq.\ (\ref{eq:zero-energy-states2}), which is resultantly also regular at the origin and behaving $\sim r^{1/2}$ when $r\rightarrow0$. Finally, an identical evaluation for the trivial regime $MB<0$ shows that there are no square integrable combinations that are also regular at the origin. This confirms that the modes are intricately related to the non-trivial $\mathbb{Z}_{2}$ nature of the bulk system. Indeed, although the above entails a particular regularization procedure, one can show more generally that, using a standard prescription 
to ensure that one chooses the right Hilbert space (i.e. domain of functions) on which the Hamiltonian acts as a self-adjoint or Hermitian operator \cite{Weidmann1987, Thaller1992,Fulop2007, Jackiw1991} ,  the obtained solution in fact pertains to the  only possible extension which allows for the existence of a zero energy state that is both localized and regular at the origin \cite{Mesaros2012b}.

\subsection{Quantum numbers of the $\pi$ flux mode}
As a sidestep, we like to remark that the above derived zero-energy modes can be associated with the same spin-charge separated quantum numbers appertaining to the famous Jackiw-Rebbi solitons in the $1+1$ dimensional Su-Schieffer-Heeger (SSH) model  \cite{Su1979, Su1983, Heeger1988}. In this regard one can view the above results as a generalization to two spatial dimensions. To make this concrete, 
reconsider the continuum $4\times 4$ Hamiltonian (\ref{eq:cont-ham}) coupled to the $U(1)$ vector potential 
\begin{equation}
H=i\gamma_0\gamma_i(k_i+A_i)+(M-B({\bf k}+{\bf A})^2)\gamma_0
\end{equation}
where the vector potential ${\bf A}$ given by Eq.\ (\ref{eq:A}). We note that the unitary matrices $\gamma_3=\sigma_2\otimes\tau_2$ and $\gamma_5=\sigma_2\otimes\tau_1$ anti-commute with the gamma-matrices $\gamma_\alpha$, $\alpha=0,1,2$. Therefore, the Hamiltonian anticommutes with the matrices $\Gamma_3\equiv i\gamma_0\gamma_3$ and $\Gamma_5\equiv i\gamma_0\gamma_5$ which then generate chiral (spectral) symmetry relating states with positive and negative energies, i.e., if $H|E\rangle=E|E\rangle$, then, for instance, $\Gamma_5|E\rangle=|-E\rangle$, and the matrix $\Gamma_5$ reduces in the zero-energy subspace of the Hamiltonian.
As a result, one can show that the  quantum numbers $\langle Q\rangle$ relative to vacuum are expressed as
\begin{equation}
\langle Q\rangle=\frac{1}{2}\left(\sum_{{\rm
occupied}}\Psi^{\top}_{E}Q\Psi_{E} -\sum_{{\rm
unoccupied}}\Psi^{\top}_{E}Q\Psi_{E}\right),
\end{equation}
where the sum has the interpretation of an integral when the spectrum is
continuous and $\{\Psi_{E}\}$ is the complete set of eigenstates of
the Hamiltonian with energy $E$.
This reveals the same spin charge separated ground state quantum numbers $\langle Q\rangle$ as those characterizing the SSH case. Namely, when both states are occupied or empty the charge is $+e$ or $-e$ and the spin quantum number is zero, whereas single occupied modes are characterized by a spin quantum number $-1/2$ or $1/2$ and a net charge zero. These results can additionally be understood from a physical perspective due to Qi and Zhang \cite{Qi2008scs}. Since a $\pi$-flux is equivalent to a $\pi$ flux, there are four different adiabatic processes that result in the same final flux configuration. That is, for each spin projection $\uparrow(\downarrow)$ the flux can adiabatically be increased from 0 to $\pm\pi$. Considering a loop around the vortex, Faraday's law of induction then implies that during such an adiabatic process a tangential electric field is induced along the loop. Accordingly, the quantized Hall conductivity per sub block thus results in a net charge transfer per spin component along the loop that simply amounts to $\Delta Q_{\uparrow(\downarrow)}=\mp e/2$ for the $\pm\pi$-flux process. Combining the four different scenarios $\phi_{\uparrow,\downarrow}=\pm\pi$, we then readily obtain the net relative charge $\Delta Q=Q_{\uparrow}+Q_{\downarrow}$ and spin $\Delta S_{z}=Q_{\uparrow}-Q_{\downarrow}$ for each case, culminating in the outlined quantum numbers.

\subsection{$\pi$ fluxes as a bulk classification probes}
Although the discrepancy in spatial dimensions allows for the spin-charge separated quantum numbers of the vortex zero-modes, we notice that their characterization is nevertheless rather reminiscent of the edge state description \cite{Koning2008}. As edge states provide for the main signature of the topological bulk entity via the bulk boundary correspondence, it therefore of topical interest to determine whether  $\pi$ flux modes can also serve as such topological indicators on a generic footing. It turns out that the answer to this question is indeed affirmative. In fact, as conducting surfaces can also pertain to trivial phases in presence of weak interactions, the characterization of a $2+1$ dimensional topological band insulator by the criterion whether a charge-0 time-reversal symmetric $\pi$ flux mode is
a Kramers doublet is actually more accurate \cite{Qi2008scs, Ran2008scs,Wen2016}.

Concerning the topological stability of the mid gap states, being the stability under continuous deformations of the vector potential, it is well-known that both the Dirac and Schr\" odinger Hamiltonian are characterized by an index theorem \cite{Aharonov1979, Jackiw1986}. However, as the continuum theory (\ref{eq:cont-ham}) entails a combination of the two, a definite index theorem mathematically relating the number of zero-modes to the non-triviality of the defining differential operator is yet to be formally established. Moreover, it is evident that in the above treatment the description of the zero-modes coincides with the manifest chiral symmetry of the continuum BHZ model \eqref{eq:cont-ham}, which in the generic context amounts to an artifact rather than a fundamental symmetry. Fortunately, the generality of the Laughlin argument allows nonetheless for an explicit verification of the intricate relation between the $\pi$-flux modes and the QSH phase also in absence of the additional spin  $U(1)_{s}$ symmetry.

As a first step it is important to realize that actually only TRS is needed to define spin-charge separated modes, as the time reversal operator $T$ acts differently on integer and half-integer spin states \cite{Qi2008scs, Ran2008scs}. Specifically, denoting the electron number operator as $N$, one can generalize the concept of spinons to arbitrary quantum states satisfying $(-1)^N|\psi_{s}\rangle=|\psi_{s}\rangle$ and $T^{2}|\psi_{s}\rangle=-|\psi_{s}\rangle$. Correspondingly, states transforming as $(-1)^{N}|\psi_{c}\rangle=-|\psi_{c}\rangle$, $T^{2}|\psi_{c}\rangle=|\psi_{c}\rangle$ define chargeons and holons. These concepts subsequently allow for the construction of an explicit invariant that quantifies the non-trivial $\mathbb{Z}_{2}$ order of the assumed unique ground state.  Namely, consider the adiabatic process of $\pi$-flux treading represented by the matrix $\Gamma$, as function of the interpolating parameter $\theta\in[0,\pi]$. In the case of a lattice model the hoppings along some cut ending on the vortex are then altered as $t_{ab}\rightarrow t_{ab}e^{i\theta(t)\Gamma}$, whereas in the continuum setting this amounts to an evident redefinition of the gauge potential. 
Crucially, for \textit{any} time reversal odd matrix $\Gamma$, satisfying $(-1)^{i\pi\Gamma}=-1$, one can then show that such an adiabatic process  results in the pumping of an integer $N_{\Gamma}$ amount of charge to the flux tube, proving that $(-1)^{N_{\Gamma}}$ defines a $\mathbb{Z}_{2}$ invariant \cite{Qi2008scs, Essin2007}. 

We note that these arguments find their foundation in the Laughlin argument. In particular, as the resulting state after spin flux threading can similarly be obtained by charge flux threading we can consider the closed path $l=l_{c}^{-1}l_{\Gamma}$ in parameter space \cite{Essin2007}. Here $l_{c},l_{\Gamma}$ refer to adiabatic paths associated with charge and spin flux threading, respectively. As the ground state is unique and the response during the charge threading process equates to zero by virtue of time reversal symmetry, we immediately conclude that an integer amount of charge is pumped to the flux tube by the spin flux threading process. Moreover, for two different adiabatic processes $\Gamma_{1}$ and $\Gamma_{2}$, the closed path $l_{\Gamma_{2}}^{-1}l_{\Gamma_{1}}$ has to result in the transport of an even multiple of charge, as it must map the unique ground state (e.g. a Kramers singlet) to itself (again a Kramers singlet ). This shows that the invariant $(-1)^{N_{\Gamma_{1,2}}}$ is independent of $\Gamma_{1,2}$. 

As a result, we see that the adiabatic $\pi$ flux threading is thus directly related to a concrete invariant characterizing the $\mathbb{Z}_{2}$ phase. Only in the case of a non-trivial $\mathbb{Z}_{2}$ invariant, an odd multiple of charge is transported to the flux tube for the appropriate $\Gamma$. Moreover, using the explicit occupation relation between the chargeons/holons and spinons in conjunction with Kramers theorem, we conclude that the {\it formation} of the mid gap modes bound to the $\pi$-flux vortex is in direct correspondence with the topological non-triviality of the bulk system.
In the specific context of the previous subsection, we may, for example, consider the process associated with  $\phi_{\uparrow}=-\phi_{\downarrow}=\pi$ represented by the matrix $\Gamma=\sigma_{z}\otimes\tau_{0}$, creating the chargeon state in the non-trivial regime. Due to the outlined arguments the associated mid gap modes are then stable to perturbations that do not close the gap, although in the absence of spectral symmetry they are no longer pinned at zero-energy.

\subsection{Dislocations as probes beyond the tenfold way \label{Dislocations2D}}
As shown above, $\pi$ flux modes represent universal topological observables that intricately relate to the $\mathbb{Z}_{2}$ topological status of the bulk. However, the derived notions come  to live even more in the context of dislocations as they can act as effective fluxes under specified conditions. 
In particular, as the piercing of $\pi$-fluxes through elementary plaquettes of the size of the lattice constant is far from experimentally viable, ubiquitous experimental defects that effectively amount to flux probes are obviously consequential in this regard.
Most importantly, due to the outlined connection to the transitional symmetry breaking of the underlying lattice, these results in turn expose an additional classification scheme that incorporates lattice symmetries beyond the tenfold way that is solely based on time-reversal symmetry  and particle-hole symmetry.

\begin{figure*}
    \includegraphics[width=0.9\textwidth]{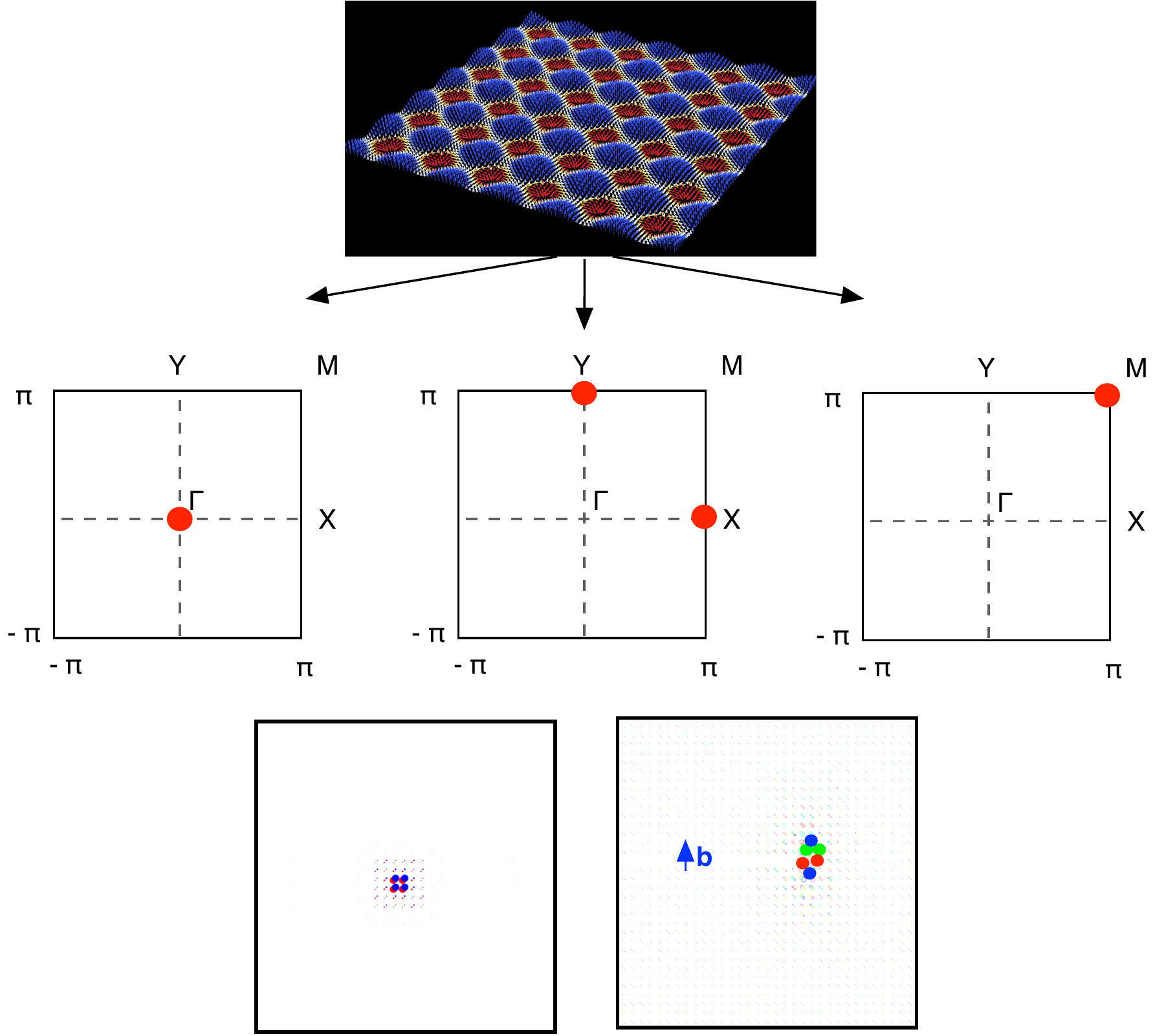}
    \caption{\label{fig::Skyr} Schematic illustration of the indexing and response to the bulk defects in case of a simple square lattice. When the $z$ component of the spin is conserved, as for the lattice regularized BHZ model \cite{Bernevig2006}, each spin projected block features a nontrivial Chern number. The resultant $\mathbb{Z}_{2}$ number is then given by the difference of these two spin Chern numbers modulo 2. As function of nearest and next nearest neighbor coupling terms one can retrieve three phases: a $\Gamma$ phase, a $X-Y$ valley phase and a $M$ phase (middle row).  These phases can be visualized by the Skyrmion lattice in the extended zone scheme (top row) for each projected spin block. That is, for each of the phases, the according points act as the loci of the Skyrmions (red dots). More generally, these momentum points then locate an opposite sign in the pfaffian structure of Equations \eqref{eq::w} and \eqref{eq::nu} and are stable to perturbative terms as long as the bulk band gap remains finite. Most importantly, the retrieved phases can also be distinguished physically upon their response to $\pi$ fluxes and dislocations. Specifically, the $\Gamma$ and $M$ phase feature a nontrivial TRS $\mathbb{Z}_{2}$ invariant and therefore bind an odd number of localized Kramers pairs (bottom row left). Here, the radii of the circles indicate the weight of the real space wave function, whereas the colors indicate the phase. In contrast only the $M$ and $X-Y$ valley phase feature dislocation modes (bottom row right), where in the latter case single pairs of modes are associated with the $X$ or $Y$ point depending whether the Burgers vector $\mathbf{b}$ entails $\mathbf{e}_{x}$ or $\mathbf{e}_y$. Hence, taking into account the response to these bulk monodromy defects each phase can uniquely be characterized. 
 }
\end{figure*}

 The simple {\it lattice regularized} BHZ model \cite{Bernevig2006} already gives away a generic wisdom in this regard.  Depending on its parameters, this model host {\it two} topological phases which are in a thermodynamic sense distinguishable: their topological nature is characterized by a Berry phase skyrmion lattice (SL) in the extended Brillouin zone (BZ), where the sites of this lattice coincide with the reciprocal lattice vectors ( which we refer to as the "$\Gamma$-phase") or with the TRS points $(\pi,\pi)$ (which we refer to as the "M-phase").
Concretely, each spin projected sub block can be rewritten in the compact form
\begin{equation}\label{eq:upper-ham}
H(k_{x},k_{y})=\mathbf{d}(\mathbf{k})\cdot\boldsymbol{\tau}, 
\end{equation}
where the matrices $\tau$ refers to the standard Pauli matrices, as in the continuum case given by Eq. \eqref{eq:cont-ham}, and $\mathbf{d}(\mathbf{k})=(\sin(k_{x}), \sin(k_{y}), M-2B[2-\cos(k_{x})-\cos(k_{y})])$. We note explicitly that the model indeed features the above mentioned two phases, which are separated by a phase transition at $M/B$=4. That is, for $0<M/B<4$ the systems is in the $\Gamma$ phase, whereas for $4<M/B<8$ the system is in the $M$ phase.
Subsequently, one can show (Fig. \ref{fig::Skyr})  that in the topologically non-trivial $\Gamma$-phase the band-structure vector field $\hat{{\bf d}}({\bf k})\equiv {\bf d}({\bf k})/|{\bf d}({\bf k})|$ forms a skyrmion centered around the $\Gamma$-point in the BZ, with corresponding skyrmion density $s({\bf k})\equiv \hat{{\bf d}}({\bf k})\cdot(\partial_{k_x}\hat{{\bf d}}({\bf k})\times\partial_{k_y}\hat{{\bf d}}({\bf k}))$. Here $s({\bf k})$ tracks the position of minimal band gap in the BZ, coinciding with it where ${\hat{\bf d}}({\bf k}) || {\partial_{ k_x}} {\hat{\bf d}}({\bf k})\times  {\partial_{ k_y}} {\hat{\bf d}}({\bf k})$.  
In the 2D extended BZ, this skyrmion structure forms a lattice which respects point group symmetry of the original square lattice. Moreover, in the $M$-phase, the skyrmion is centered at the $M$ point in the BZ. The position of the corresponding
skyrmion lattice relative to the extended BZ is therefore different than in the $\Gamma$-phase, although the skyrmion lattice still respects the full point group symmetry of the square lattice. On the other hand, in the topologically trivial phase, the vector field ${\bf d}$ forms no skyrmion in the BZ, consistent with the vanishing of topologically invariant spin Hall conductance $\sigma^S_{xy}=(4\pi)^{-1}\int_{BZ}d^2{\bf k}\;s({\bf k})$. The position of the skyrmion lattice relative to the extended BZ thus encodes translationally active topological order which, similar to the above ideas, is probed by the lattice dislocations.   

Let us accordingly first motivate analytically, using elastic continuum theory, that a lattice dislocation should effectively amount to the above magnetic $\pi$-flux problem {\it in the $M$ phase  only}, thereby signaling the correspondence to the translational symmetry. To this end we consider a dislocation with Burgers vector ${\bf b}$, and expand the Hamiltonian (\ref{eq:upper-ham}) around the M-point in the BZ. As a next step, the dislocation introduces an elastic deformation of the medium described by the distortion $\{{\bm \varepsilon}_i({\bf r})\}$ of the (global) Cartesian basis $\{{\bf e}_i\}$, $i=x,y$, in the tangent space at the point ${\bf r}$ \cite{Landau1981ela}. Consequently, the momentum in the vicinity of the $M$-point reads 
\begin{equation}
k_i={\bf E}_i\cdot({\bf k}_M-{\bf q})=({\bf e}_i+{\bm \varepsilon}_i)\cdot({\bf k}_M-{\bf q}),
\end{equation} 
where ${\bf k}_M=(\pi,\pi)/a$, ${\bf q}$ is the momentum of the low-energy excitations, $|{\bf q}|\ll |{\bf k}_M|$, and we have restored the lattice constant $a$. 
The corresponding continuum Hamiltonian after this coarse graining \cite{Bausch1998} then assumes the form
\begin{equation}\label{eq:cont-ham-M}
H_{\rm eff}({\bf k},{\bf A})=\tau_i(k_i+A_i)+[{\tilde M}-{\tilde B}({\bf k}+{\bf A})^2]\tau_3,
\end{equation}
in terms of the redefinitions ${\bf q}\rightarrow{\bf k}$, ${\tilde M}\equiv M-8B$, ${\tilde B}\equiv-B$, and $A_i\equiv-{\bm \varepsilon}_i\cdot{\bf k}_M$. The form of the distortion ${\bm \varepsilon}_i$ is determined from the dual basis in the tangent space at the point ${\bf r}$ which in the case of a dislocation with ${\bf b}=a {\bf e}_x$ entails ${\bf E}^x=(1-\frac{ay}{2\pi r^2} ){\bf e}^x+\frac{ax}{2\pi r^2}{\bf e}^y, {\bf E}^y={\bf e}^y$ \cite{kleinert}. Using that ${\bf E}_i\cdot{\bf E}^j=\delta_i^j$, we obtain the distortion field to the leading order in $a/r$,  ${\bm\varepsilon}_x=\frac{ay}{2\pi r^2}{\bf e}_x,{\bm\varepsilon}_y=-\frac{ax}{2\pi r^2}{\bf e}_y $.
This form of the distortion yields the vector potential 
\begin{equation}\label{eq:pi-flux}
{\bf A}=\frac{-y{\bf e}_x+x{\bf e}_y}{2r^2}
\end{equation}
in Eq.\ (\ref{eq:cont-ham-M}), demonstrating that the dislocation in the $M$-phase acts as a magnetic $\pi$-flux. On the other hand, in the $\Gamma$-phase, the continuum Hamiltonian has the generic form (\ref{eq:cont-ham-M}), with ${\tilde M}=M$ and ${\tilde B}=B$. However, the action of the dislocation in this case is trivial, since the band gap is located at zero momentum rendering ${\bf A}=0$. Applying the above results, we therefore conclude that a dislocation must bind a Kramers pair mid gap modes in the $M$-phase, while having no effect in the $\Gamma$-phase.  As a result, dislocations act as the universal probe distinguishing the two non-trivial phases carrying the same TRS invariant.

These notions can readily be numerically corroborated  \cite{Juricic2012, Mesaros2012b} by performing a numerical analysis on the tight-binding BHZ model \cite{Bernevig2006} in real space 
\begin{equation}
  \label{eq:1}
  H_{TB}=\sum_{{\bf R},{\bm \delta}}\Bigg(\Psi_{\bf R}^\dagger\left[ T_{\bm \delta}+i\frac{R_0}{2}(\tau_{0}+\tau_3){\bf e}_z\cdot({\bm \sigma}\times {\bm \delta})\right]\Psi_{{\bf R}+{\bm \delta}}+\Psi_{\bf R}^\dagger\frac{\epsilon}{2}\Psi_{\bf R}+H.c.\Bigg).
\end{equation}
Here, $\Psi_{\bf R}=(s_\uparrow({\bf R}),p_\uparrow({\bf R}),s_\downarrow({\bf R}),p^*_\downarrow({\bf R}))$ annihilates the $\sim|s\rangle$ type, and $\sim|p_x+i p_y\rangle$ type orbitals at site ${\bf R}$ and nearest neighbors ${\bm \delta}\in\{{\bf e}_x,{\bf e}_y\}$. Furthermore, we have defined 
\begin{equation*}
T_{{\bm\delta},\uparrow\uparrow}=
\begin{pmatrix}
\Delta_{s}&t_{\bm \delta}/2\\
t'_{\bm \delta}/2&\Delta_p
\end{pmatrix}
\end{equation*}
 and $T_{{\bm \delta},\downarrow\downarrow}=T^*_{{\bm \delta},\uparrow\uparrow}$ in terms of $t_x=t'_x=-i$, $t_y=-t'_y=-1$, $\Delta_{s/p}=\pm B$ and on-site energies $\epsilon=[(M-4B)\tau_3]\otimes\sigma_0$. Finally, the $R_0$ term is the nearest neighbor Rashba spin-orbit coupling \cite{Rothe2010} which is induces a breaking of the $z\rightarrow -z$  reflection symmetry and resultantly the PHS symmetry. Taking into account various shapes, systems sizes, finite Rashba coupling and even disorder, one can then verify that the dislocations modes persist in the $M$ phase as long as the gap is maintained \cite{Juricic2012, Mesaros2012b}. As a result, we thus conclude that dislocations indeed signal an additional classification scheme refining the ten fold way. Moreover, these findings can also be considered consequential in the view of experimental verification. Dislocations are ubiquitous in any real crystal, for
instance in the form of small angle grain boundaries, and one thus
expects that their cores should carry spin charge separated dislocation modes. These modes
should detectable with scanning tunneling spectroscopy.
Nonetheless, the real experimental challenge arguably lies in the realization
of non-$\Gamma$ TIs that are also easily accessible in the light of 
spectroscopic measurements.

\section{Connecting the dots\label{sec::Classification}}
As alluded to, the dislocation results do not  stand by themselves but in fact instigate a physical indexing  \cite{Slager2013}. After all, the previously described $\Gamma$ and $M$ phase can be distinguished upon their response to dislocations, whereas the nontrivial TRS invariant is justly probed by a $\pi$ flux  in both instances. The according appearing classification scheme, which goes hand in hand with these results, then allows to 
treat the two and three dimensional case on a generic footing and reduces to the familiar notions, such as weak indices, when one only time reversal symmetry is taken into account.

As a starting point to make this concrete one may depart from the general construction, Eqs. \eqref{eq::w} and \eqref{eq::nu}, of Fu and Kane \cite{Fu2007a, Fu2007b,Fu2006}.
As a next step, the generality of Eq. (\ref{eq::w}) in conjunction with its momentum dependence can then naturally be exploited to signify the natural role of the underlying crystal symmetries.
In this regard, firstly notice that the set of ${\Gamma}_i$ points at which the Hamiltonian commutes with the time-reversal operator is in fact set by the space group of the lattice. Secondly, it is evident that we may choose  the overall phase of the Bloch wavefunctions so that a unique phase, which we dub the ``$\Gamma$'' phase, has $\delta_{{\Gamma}}=-1$ at the ${\Gamma}$-point in the BZ and $\delta_i=1$ at all the other high symmetry points. The crucial observation then pertains to the fact that \emph{the distribution} of signs of the Pfaffian, $\delta_i$, at the points $\Gamma_i$, and not only their product, encodes for additional structure. To show this, we first consider how the matrix of overlaps transforms under a lattice symmetry operation represented by a unitary operator $U$
\begin{equation}\label{eq::symmetryrelation}
w_{mn}(\mathbf{k})=\langle u_{m}(-\mathbf{k})|\vartheta | u_{n}(\mathbf{k})\rangle=\langle u_{m}(-U\mathbf{k})|U\vartheta U^{\dagger} | u_{n}(U\mathbf{k})\rangle=w_{mn}(U\mathbf{k}).
\end{equation}
As a consequence, when some of these high symmetry points are related by a point-group symmetry of the lattice, the signs of their Pfaffian expression $\delta_i$  have to be equal.
Therefore, it is sufficient to consider a subset, $\Gamma_a$, of representative, inequivalent high symmetry points that are also \emph{not related} by any symmetry.
This leads to the following rule that allows for determination of all the topological phases given the space group and the corresponding high symmetry points, ${\Gamma}_i$: each phase is obtained by selecting a single representative high-symmetry point $\Gamma_a$ and setting $\delta_{{\Gamma_a}}=-1$ , which automatically sets $\delta_{{\Gamma_b}}=-1$ at all the high-symmetry points $\Gamma_b$ related by point group symmetry to $\Gamma_a$. We emphasize that all these signs can be reversed, rendering the distribution unaffected. The different phases are then separated by topological quantum phase transitions that involve a bulk band gap closing which then changes the values of $\delta_i$'s.

This simple classification principle can be illustrated by reconsidering the different topological phases on the square lattice (Figure \ref{fig::Skyr}). Namely, the $\Gamma$ phase, with the band inversion at the $\Gamma$ point, is characterized by  $\delta_{{\Gamma}}=-1$, and $\delta_{{X}}=\delta_{{Y}}=\delta_{{M}}=1$ in terms of  the time-reversal invariant (TRI) momenta $X$, $Y$, and $M$ in the BZ. Similarly, the $M$ phase, with the band inversion at the $M$ point, is signified by  $\delta_{{M}}=-1$, and $\delta_{{\Gamma}}=\delta_{{X}}=\delta_{{Y}}=1$. Furthermore, we also observe that, since the $X$ and the $Y$ points are related by a $C_4$ rotation, there can exist one other phase having $\delta_{X}=\delta_Y=-1$, and $\delta_\Gamma=\delta_M=1$.
The product of the $\delta_i$'s at all TRI momenta yields in this case a trivial $Z_2$ invariant, $\nu=0$. However,  it turns out that the $C_4$ rotational symmetry protects this phase, since it pins the band inversions at the ${X}$ and ${Y}$ points, and therefore represents a ``valley'' or ``crystalline''\cite{Fu2011} insulator -- a phase trivial tenfold way-wise but protected by the lattice symmetries. Interestingly, this phase can indeed be found within the context of a simple model. That is, when one supplements the BHZ model with next nearest neighbor hopping terms, a phase with $\delta_{X}=\delta_Y=-1$ and $\delta_\Gamma=\delta_M=1$ can be retrieved as function of the model parameters in addition to the $\Gamma$ and $M$ phase \cite{Slager2013}. As anticipated, this phase hosts now two pairs of counter propagating edge states. Moreover, evaluating the response to $\pi$ fluxes and dislocations, it can once again be distinguished. That is, introducing a $\pi$ flux does not results in an odd pair of modes. Nonetheless, as the opposite sign in the $\delta_i$ configuration is located at the $X$ and $Y$ point, inserting a dislocation does result in a single pair of modes as, depending on the orientation of the Burgers vector $\mathbf{b}$, the $\mathbf{k}\cdot\mathbf{b}$ argument amounts to a phase $\pi$ for only one of them. Moreover, within this context, breaking of the $C_4$ symmetry consistently leads to instability of this phase \cite{Slager2013}.

\begin{figure}[h]
\includegraphics[scale=0.16]{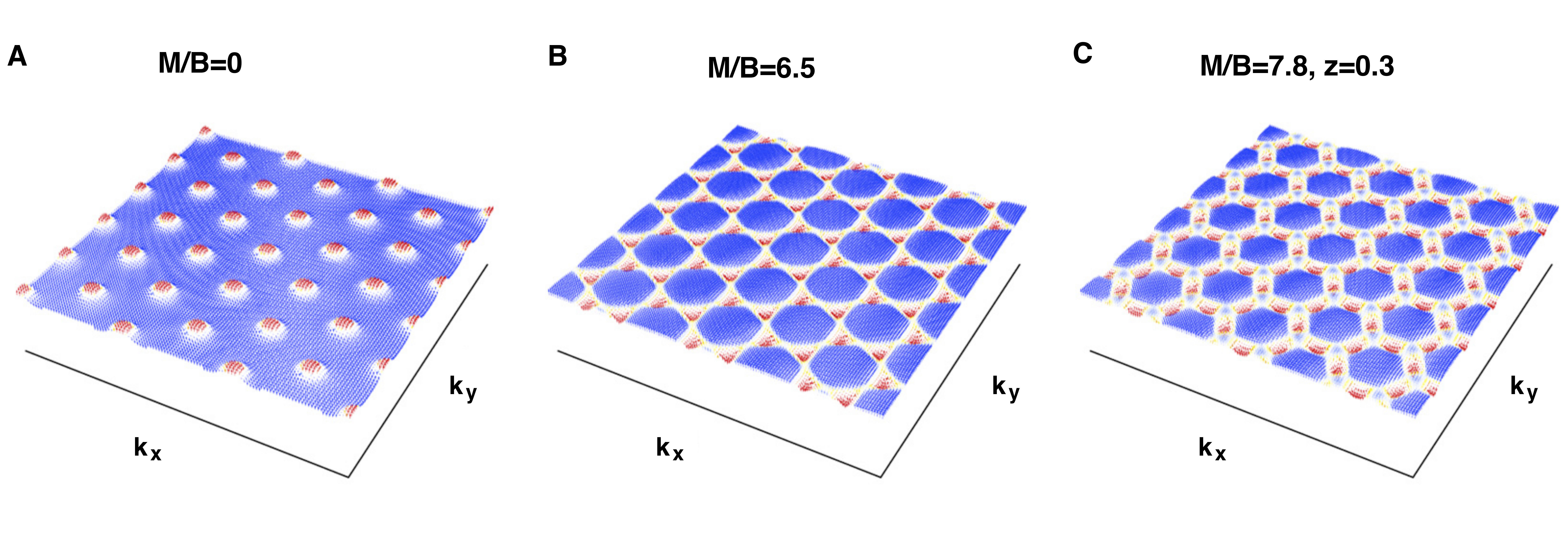}
\caption{\label{fig::hex}The skyrmion lattices in the extended Brillouin zone for the three topological phases on the primitive hexagonal (triangular) lattice for the spin up projection (adopted from \cite{Slager2013}).  Specifically, the phases are retrieved within the BHZ model setting \cite{Bernevig2006} as function of the ratio $M/B$ of nearest neighbor hopping parameters  and a similar ratio $z$ of next nearest neighbor hopping parameters. The loci of the Skyrmions are colored in red and denote the phase. ({\bf A}) The Skyrmion lattice for the $\Gamma$ phase. ({\bf B})  The Skyrmion lattice for the $K$ valley phase. ({\bf C}) When next nearest neighbor hopping is taken into account the system can enter the $M$ phase, which has the Skyrmions positioned at the $M$ points. The presence of all three phases within this simple model elucidates that besides the usual TRI points also the corners of the Brillouin zone act as TRI points by virtue of the sixfold rotational symmetry. }
\end{figure}

As further impetus to the confirmation of the relevance of the above, we remark that the set of TRI points can already be extended in the above setup on a triangular lattice (Figure \ref{fig::hex}). 
Concretely, one notices that not only the usual four momenta, comprising the $\Gamma$ point and the midpoints of the edges of the hexagonal BZ, but also the corners entail TRI points. The latter is due to the effect of the sixfold $C_6$ symmetry, which ensures that the Hamiltonian at these momenta $\mathbf{K}$ equate to $-\mathbf{K}$. In the above spirit one then expects three representatives $\Gamma_a$ and thus three phases. Formulating the BHZ model, including next nearest neighbor terms, on the triangular lattice and taking a specific spin projection, these three phases can indeed be retrieved and accordingly visualized by their Skyrmion lattices in the above spirit \cite{Slager2013}.

As a result, the above rule allows us to consistently index the topological phases. The set of BZ high-symmetry points $\Gamma_i$ at which there is band inversion, i.e., $\delta_{\Gamma_i}=-1$, is invariant under the operations of a subgroup of the lattice space group. This symmetry subgroup therefore protects, and labels, the topological phase. The other element in this indexing is the protection by TRS (T), existing when the $Z_2$ invariant $\nu=1$, giving, for instance $T-p4mm$ as the $\Gamma$ phase on the square lattice. When the protecting symmetries coincide between phases, we explicitly label $\Gamma_i$ (lower index), as, e.g.,  for $T-p2m_X$, $T-p2m_Y$ and $T-p2m_M$ phases on the rectangular lattice.
This leads to the list of topological phases in 2D presented in Table \ref{2dTable} conveying 18 distinct topological phases.
As a general {\it main result}, one thus obtains two additional broad classes of topological states protected by TRS or crystalline symmetries,
besides the class of states robust against general TRS perturbations ($\Gamma$-states): translationally-active states protected both by TRS and lattice symmetry, responding to dislocations, and topological crystalline insulators which are tenfold-way-wise trivial but protected by space group symmetry and also susceptible to dislocations \cite{Fu2011}.
\begin{table}
\centering
\footnotesize
\begin{tabular}{c|c|c|c|c}
\hline
\hline
{ Bravais Lattice (PG)} & { WpG}& ${\Gamma}_{i}$ &{$\delta_i$}&
Index (Phase)  \\
\hline
{Square}{ ($D_{4}$)} & $p4mm$ & (${\Gamma}$,
${M}$, ${X}$, ${Y}$) &(-1,1,1,1) &$T\text{-}p4mm$
($\Gamma$)\\ \cline{4-5}
&$p4gm$ &  &(1,-1,1,1) &$T\text{-}p4$ ($M$)\\ \cline{4-5}
 &$p4$ & &(1,1,-1,-1) &$p4$ ($X$-$Y$-valley)\\ \cline{4-5}
 \hline
 {Rectangular}{ ($D_{2}$)} & $p2mm$ & (${\Gamma}$,
${M}$, ${X}$, ${Y})$&(-1,1,1,1)
&$T\text{-}p2mm$ ($\Gamma)$\\ \cline{4-5}
& $p2mg$  & &(1,-1,1,1)&$T\text{-}p2m_{M}$ ($M$)\\ \cline{4-5}
&  $p2gg$ & &(1,1,-1,1) &$T\text{-}p2m_{X}$ ($X$)\\ \cline{4-5}
& $pm$, $pg$  & &(1,1,1,-1) &$T\text{-}p2m_{Y}$ ($Y$)\\ \cline{4-5}
\hline
{Rhombic}{ ($D_{2}$)} & $c2mm$ & $({\Gamma}$,
${M}_{0}$, ${M}_{-1}$, ${M}_{1})$&(-1,1,1,1) &$T\text{-}c2mm$ ($\Gamma$)\\ \cline{4-5}
&$cm$  & &(1,-1,1,1) &$T\text{-}c2m$ ($M_0$)\\ \cline{4-5}
& & &(1,1,-1,-1)&$c2m$ ($M$-valley)\\ \cline{4-5}
\hline
{Oblique}{ ($C_{2}$)} &$p2$&$({\Gamma}$,
${M}_{0}$, ${M}_{-1}$, ${M}_{1})$ &(-1,1,1,1)
&$T\text{-}p2$ ($\Gamma$)\\ \cline{4-5}
&$p1$  &&(1,-1,1,1)&$T\text{-}p2_{M_{0}}$ ($M_0$)\\ \cline{4-5}
& & &(1,1,-1,1) &$T\text{-}p2_{M_{\text{-}1}}$ ($M_{-1}$)\\ \cline{4-5}
& & &(1,1,1,-1)&$T\text{-}p2_{M_{1}}$ ($M_1$) \\ \cline{4-5}
\hline
{Hexagonal} & $p6mm$ & $({\Gamma}$,
${M}_{0}$, ${M}_{-1}$, ${M}_{1}$,
${K}_{-}$, ${K}_{+}$)&(-1,1,1,1,1,1) &$T\text{-}p6mm$ ($\Gamma$)\\ \cline{4-5}
 {\it(hexagonal -- $D_{6}$)}&$p6$ & &(1,-1,-1,-1,1,1) &$T\text{-}p6$ ($M$)\\ \cline{4-5}
& & &(1,1,1,1,-1,-1)&$p6$ ($K$-valley)\\ \cline{2-5}
Hexagonal& $p3m1$&$({\Gamma}$,
${M}_{0}$, ${M}_{-1}$, ${M}_{1}$)&(-1,1,1,1)&$T\text{-}p3m1$ ($\Gamma$)\\
 {\it(rhombohedral -- $D_{3}$)} & $p31m$, $p3$  & &&\\
\hline
\hline
\end{tabular}
\caption{ Table of the topological phases in two dimensions \cite{Slager2013}. For each of the lattice structures, the corresponding point-group (PG) symmetry and the possible wallpaper groups (WpG), i.e. space group, are given. The corners of the square and rectangle are denoted by $M$, whereas in the triangular Bravais structure they are indicated by $K$. Additionally, the centers of the edges are denoted by $X$ and $Y$ in both the square and rectangular case and by $M$ in the other lattices \cite{dresselhausbook}. The resulting phases are characterized by the distribution of $\delta_i$ at the $\Gamma_i$ points consistent with the WpG symmetry. The possible phases have been clustered according to the Bravais lattices, with the hexagonal structure being the only exception. In this case the WpGs containing six-fold and three-fold rotational symmetries relate the high symmetry points in different ways. As a result, the Hamiltonian does not commute with the time-reversal operator at the $K$ points in the latter case. The obtained phases are ultimately protected by TRS  (whenever $\nu=1$), WpG symmetry, or both, and are accordingly indexed. The index (last column) describes the part of wallpaper group that leaves the subset $\Gamma_i$ having $\delta_{i}=-1$ invariant, while the additional label 'T' denotes TRS protection. In the column denoted ``Phase'' we introduce a convenient but imprecise shorthand notation.}
\label{2dTable}
\end{table}

\begin{figure}
\center
\includegraphics[scale=0.6]{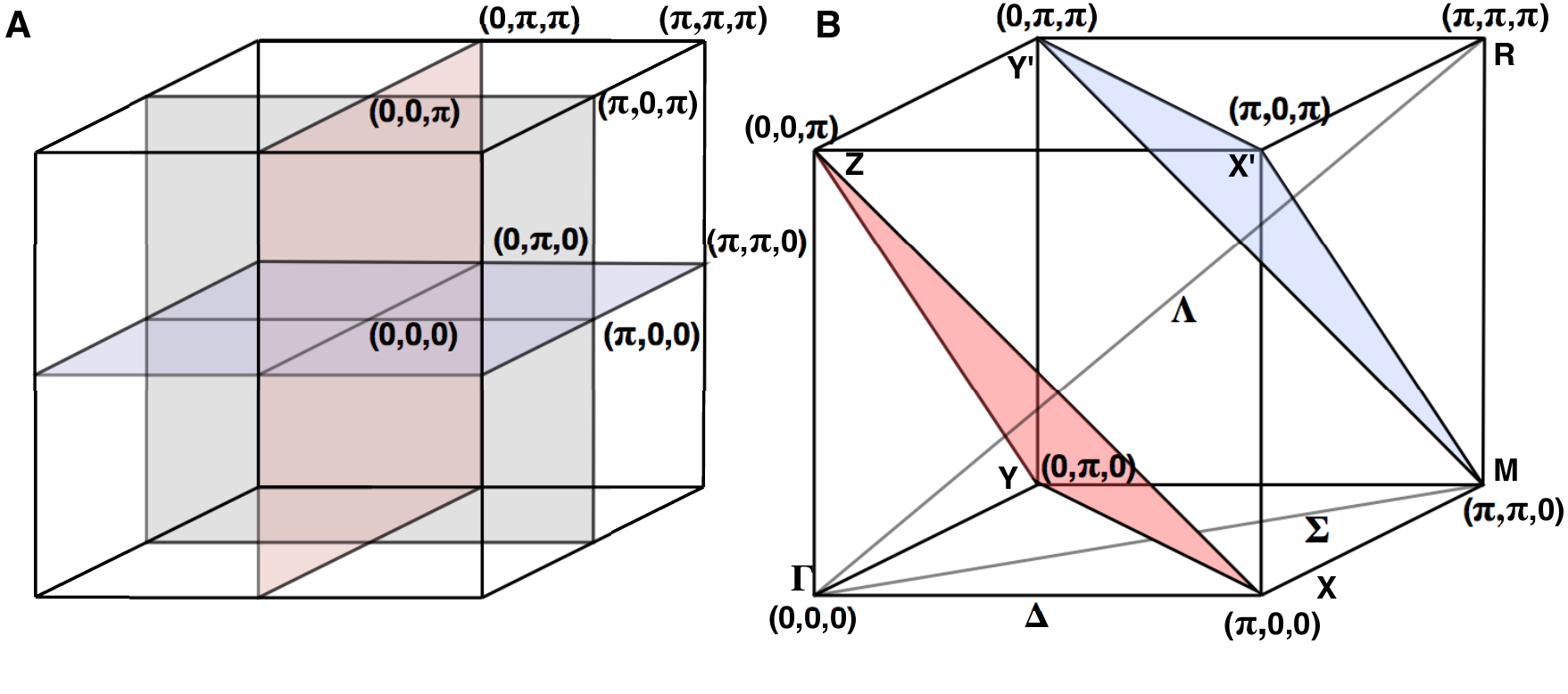}
	\caption{ Role of lattice symmetries in the classification of topological states. ({\bf A}) The eight TRI momenta in the Brillouin zone of the primitive cubic lattice. When only TRS is considered the sign of any quadruple of $\delta_{i}$'s within a plane connecting them can be changed, leaving their product the same. As a result one obtains, in addition to the 'strong' invariant $\nu_{0}$, three weak invariants corresponding to the orthogonal planes. ({\bf B}) The constraints on the ${\delta}_i$'s arising from the lattice symmetries.  The high symmetry axes  $\Delta$, $\Lambda$ and $\Sigma$
represent axes of four-, three- and two-fold rotations, respectively; these transform the TRI points in the colored planes into each other, and thus constrain the corresponding $\delta_{i}$'s to be equal. }
\label{fig:cubic}
\end{figure}

Most appealingly, the outlined evaluations can directly be applied in case of three spatial dimensions, albeit becoming more involved given the 230 space groups and the large number of possible high-symmetry points. Turning to the most readily obtained example of the primitive cubic lattice with the familiar eight TRI points (Fig. \ref{fig:cubic}), it is evident that the presence of the threefold rotations about the diagonals ensure that the $\delta_{i}$ at the momenta $(X,Y,Z)$ as well as $(X',Y',M)$ are related. Consequently, the anticipated different phases  arise by examining the sign structure of the according four representatives $\Gamma$, $X$, $X'$ and $R$. These phases can correspondingly be retrieved within the familiar setup \cite{Slager2013}. We moreover emphasize  the impact on the weak index procedure \cite{Fu2007a, Moore2007} in this regard. For instance, the $T-pm{\bar 3}m$ ($\Gamma$) and $T-p3(4)_R$ ($R$) phases correspond with $(1;0,0,0)$ and $(1;1,1,1)$ indices, respectively. The latter two indices can, however, also be associated with the $T-p3(4)_M$ and $T-p3(4)_X$ phases. More generally, we note that a true layering can only occur if the underlying crystal symmetry allows for it. Indeed, considering the cubic example, a layering in the $\hat{z}$-direction, characterized by a $(0;0,0,1)$ invariant, is not possible due to the mentioned threefold rotations. Hence, this possibility can only be attained by breaking the cubic symmetry by e.g. stretching the $\hat{z}$-direction, turning the lattice into a tetragonal form. We emphasize that the $\delta_{i}$ configuration then exactly matches the weak indices by having the different signs on the $M$ and $R$ points. 

As a final remark, we note that it is of significant interest to put these physical interpretations on a mathematical solid footing. In this regard it is interesting that the basic idea of considering how specific momenta in the Brillouin zone are linked has recently be shown to lead to a direct mapping to the underlying mathematical structure in certain scenarios. Specifically, considering class A systems, that is systems in the presence of lattice symmetries solely, the connections between representations at the high symmetry momenta translate into a set of gluing conditions \cite{Kruthoff2016}. Solving these simple combinatorial constraints then reveals all specific invariants that can directly be linked to the descriptive equivariant $K$-theory \cite{Freed2013}. Using these conditions and devising a modulo construction by dividing out by phases that are adiabatically connected to the trivial one, this then gives a handle to determine indicators of band structures in a mathematical controlled manner in the presence of additional symmetries such as TRS \cite{Po2017, Bradlyn2017}. The precise relation to the above as well as the descriptive $K$-theory description however still entails for a very  interesting future pursuit. 

\section{Three dimensional dislocations revisited\label{sec::Modes3D}}
With the above perspective in hand, it is worthwhile to revisit the study of dislocations modes along line defects \cite{Ran2009,Tteo2010, Slagerprb2015} in three spatial dimensions and reconfirm the accordance via a generalized mechanism. On first sight this may appear to pose somewhat of a conundrum as the outlined characterization essentially corresponds to a 'specific monopole configuration' in the Brillouin zone whereas dislocations in this scenario entail line-like defects. Nonetheless, taking into account the dislocation line orientation ${\bf t}$, Burgers vector ${\bf b}$ and the band-inversion momentum ${\bf K}_{\rm inv}$, it turns out that the criteria for the formation of gapless propagating modes along these line defects
can effectively and consistently  be captured by what we shall refer to as ${\bf K}\text{-}{\bf b}\text{-}{\bf t}$ rule \cite{kbt2014}.

The crucial insight to connect these notions revolves around the role of the translational symmetry along the dislocation line. In particular, if a dislocation line is, for instance, oriented along the $z$-axis (${\bf t}={\bf e}_z$) the 
lattice Hamiltonian may effectively be written as
\begin{equation}\label{eq:ham-dim-red}
H_{\rm 3D}(x,y,z)=\sum_{k_{z}}e^{ik_z z}H_{\rm eff}^{\rm 2D}(x,y,k_z).
\end{equation}
Where we note that the 2D lattice Hamiltonian $H_{\rm eff}^{\rm 2D}$ possesses the (wallpaper group) symmetry of the crystallographic plane orthogonal to the dislocation line, because the Burgers vector always entails a Bravais lattice vector. 

\begin{figure}
\center
 \includegraphics[scale=0.449]{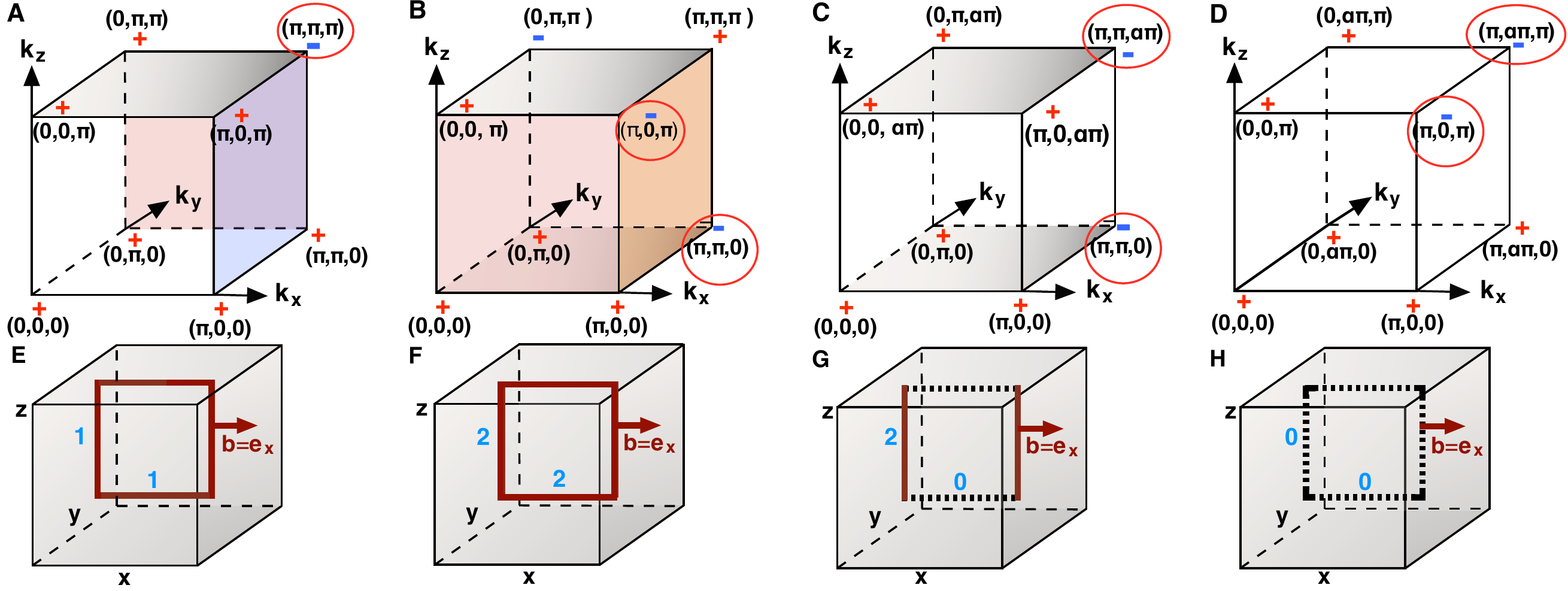}
\caption{Illustration of the ${\bf K}\text{-}{\bf b}\text{-}{\bf t}$ rule (adopted from \cite{kbt2014}). In essence this rule elucidates the interplay between the electronic topology in the momentum space (top panels),  and the effect of dislocations in real space (bottom panels).
 Panels A to D show the electronic-band topology of the  $T\text{-}p3(4)_{R}$ and $T\text{-}p3(4)_{M}$ 
 phases on a simple cubic lattice and the  $p4_{R,M}$ and $p4_{X',R}$ weak phases on tetragonal lattices \cite{Slager2013}.
 A dislocation with Burgers vector ${\mathbf b} = {\bf e}_x$  acts on the encircled TRI
momenta in the planes orthogonal to the dislocation line. As a result, the colored planes host an effective $\pi$ flux. The resulting number of Kramers pairs of helical modes along the edge and screw parts of the loop is indicated in  blue.
({\bf A}) The symmetric $T\text{-}p3(4)_{R}$ phase has a topologically non-trivial plane hosting a $\pi$ flux orthogonal to any of the three crystallographic directions and hence any dislocation loop binds modes along the entire core, as shown for a loop in the $\hat{x}-\hat{z}$ plane (see panel {\bf E}). ({\bf B}) In the $T\text{-}p3(4)_{M}$ phase, translationally active phases in the TRI planes orthogonal to $k_z$ and a valley phase in $k_x=\pi$ plane host $\pi$ fluxes. Hence the dislocation loop binds two pairs of modes, as displayed in panel {\bf F}.
These modes are symmetry-protected against mixing. ({\bf C}) In the $p4_{M,R}$ phase, only the TRI planes normal to $k_{z}$ host an effective $\pi$-flux and hence the same dislocation loop binds modes only to the edge-dislocation parts, as displayed in panel {\bf G}. These modes are not protected against mixing. ({\bf D}) In the $p4_{X',R}$ phase all TRI planes orthogonal to the dislocations lines have a trivial flux and neither the edge nor the screw dislocation of the loop binds  modes, as illustrated in panel {\bf H}.}
\label{figure::cvone}
 \end{figure}

Given this reduction procedure, one can then readily consider the effect on the electronic side of the mentioned interplay, which is characterized by the band-inversions at the time-reversal invariant (TRI) momenta  ${\bf K}_{\rm inv}$ in the Brillouin zone.  As a dislocation disturbs the crystalline order only microscopically close to its core, the continuum theory of elasticity can be used in the same manner as above, see Subsection \ref{Dislocations2D},  to  verify that the dislocations give rise to an effective flux  $\Phi={\bf K}_{\rm inv}\cdot {\bf b}$ \cite{kbt2014}. Note in particular that the corresponding $U(1)$ gauge field $A_i=-{\bm\varepsilon}_i\cdot{\bf K}_{\rm inv}$ that minimally couples to the electronic excitations only has non-trivial components in the plane orthogonal to the dislocation line, consistent with Eq.\ (\ref{eq:ham-dim-red}). Given the previous it is  directly inferred that, when this flux  $\Phi\,({\rm mod }\, 2\pi)$ is nonzero, the dislocations host propagating helical modes provided that  the 2D Hamiltonian in a TRI plane orthogonal to the dislocation line  $\hat{\bf t}\equiv{\bf t}/|{\bf t}|$ is topologically non-trivial in the exact sense of the above space group classification. That is, only when the projected Hamiltonian $H_{\rm eff}^{\rm 2D}(k)$ features an odd number of {\it non-symmetry} related translationally-active inversion momenta, mid gap states at that momentum $k$ are developed for non-zero $\Phi$. The full spectrum of modes, for each $k$, is then protected by the lattice symmetry that relates the gauged momenta ${\bf K}_{\rm inv}$, i.e those monenta giving rise to a nonzero flux. 
This is what we refer to as the ${\bf K}\text{-}{\bf b}\text{-}{\bf t}$ rule (see Fig. \ref{figure::cvone}).

This rule and the following descendant construction together imply  that the bound states for a given ${\bf k}\cdot\hat{\bf t} $ momentum combine into a spectrum
of propagating helical modes along the dislocation line. For ${\bf k}\cdot\hat{\bf t}={\bf K}_{\rm inv}\cdot\hat{\bf t}$, the system develops a Kramers pair of true zero modes $\Psi_0\equiv(\psi_0,T\psi_0)^\top$, with $T$ representing  the time-reversal operator satisfying $T^2=-1$. We should stress here that for this specific momentum the reduced system precisely reduces to a representative of the 2D problems considered in the above and that the term zero mode should be interpreted in this manner. This is a consequence of the fact that by definition the ${\bf K}_{\rm inv}$ point is a TRI point hosting a band inversion.
Deviating from this 'parent'  momentum by $ {\bf q}\cdot \hat{\bf t}$,  the effective low-energy Hamiltonian for the propagating modes then generally develops a linear gap $H_{\rm eff}\sim {\bf q}\cdot \hat{\bf t}$, to lowest order. The gapped Kramers pair of descendant states are then present as long as $H_{\rm eff}(q)$ remains in the topological non-trivial phase and may then be captured by $H_{\rm eff}=v_{ t}\Sigma_3({\bf q}\cdot \hat{\bf t})+{\mathcal O}(q^2)$. Where,  the Pauli matrix $\Sigma_3$ acts in the two-dimensional Hilbert space of the dislocation modes, which are of the form $\Psi_{{q_t}}\equiv(\psi_0\,e^{i ({\bf q}\cdot\hat{\bf t})({\bf r}\cdot\hat{\bf t})},(T\psi_0)\,e^{-i ({\bf q}\cdot\hat{\bf t})({\bf r}\cdot\hat{\bf t})})^\top$, and $v_{ t}$ is the characteristic velocity, which is set by the symmetries and details of the band structure.

The workings of these ideas can be illustrated further by examining the specific possibilities, which can all be confined by numerical means \cite{kbt2014}. Considering for example the $T\text{-}p3(4)_{R}$ (Figures \ref{figure::cvone}(A) and (E)), being translationally active and protected by TRS, we 
see that if $\mathbf{b}=\mathbf{e}_{x}$ and ${\bf t}=\mathbf{e}_{z}$, the effective Hamiltonian in the $k_{z}=\pi$ plane reduces to the $\pi$ flux problem in ${M}$ phase and hence results in a pair zero modes. Deviating away from this momentum, these modes form a dispersion as there is necessarily a mass term, depending on $k_z$, to ensure that the Hamiltonian at $k_z=0$ is trivial. In fact, by virtue of the band inversion being at at momentum $(\pi,\pi,\pi)$, we see that any combination of $\mathbf{b}$ and ${\bf t}$ results in the same analysis and therefore any line or loop binds modes. We note that TRS together with the crystal symmetry evidently protects the modes from backscattering in the latter case. Indeed, an odd number of modes cannot be gaped and should be bound to the whole defect. This consistency condition is thus naturally associated with ${\bf K}\text{-}{\bf b}\text{-}{\bf t}$ rule.  As an ultimate consequence, we observe that such a scenario in a completely isotropic lattice realizes a remarkable strong variant of this effect. Namely, as in this case all velocities in every direction must be the same, the gapless states in the dislocation loop must propagate in a way completely oblivious to the lattice directions!

As a next step, consider the system in the $T-p3(4)_{M}$ phase with the band-inversion located at the momenta $M\equiv(\pi,\pi,0)$, $X'\equiv(\pi,0,\pi)$, $Y'\equiv(0,\pi,\pi)$, which are related by the threefold rotational symmetry (Figures \ref{figure::cvone}(B) and (F)). According to the ${{\bf K}\text{-}{\bf b}}\text{-}{\bf  t}$ rule, for the edge-segment of the dislocation loop, having  $\mathbf{b}=\mathbf{e}_{x}$, both the $k_z=0$ and $k_z=\pi$ planes host an effective $\pi$-flux, originating from the $M$ and $X'$ points. Additionally, the $k_x=\pi$ plane hosts a valley/crystalline phase, and thus the screw-dislocation parts also host two pairs of modes. Therefore, there is a total of two Kramers pairs of gapless dislocation modes along the loop, which are {\it protected by symmetry}. Note that their existence crucially depends on the fact that the $M$ and $X'$ momenta are symmetry-related. Were this not the case, the weak index of this $T-p3(4)_{M}$ phase $M_i=(0,0,0)$ would consistently predict no dislocation modes at all.

One can subsequently move away from the cubic symmetry by considering the $p4_{M,R}$ phase on a tetragonal lattice with $a_x=a_y=a$, $a_z=a/\alpha$, where $a_i$ is the lattice constant in the direction ${\bf e}_i$, and $\alpha\neq1$ is the lattice deformation parameter (Figs. \ref{figure::cvone}(C) and (G)). Note here the subtle difference between the resulting weak phase and a crystalline/valley phase, which has an even number of band-inversions {\it protected} by a 3D space group symmetry. The usual strong and weak indices  \cite{Moore2007, Fu2007a, Ran2009} in this phase are $(\nu; M_i)=(0;0,0,1)$, and thus  $\mathbf{M}\cdot\mathbf{b}=0$. Nonetheless,  the $k_z=0$ and $k_z=\alpha\pi$ planes are topologically non-trivial therein (Fig. \ref{figure::cvone}). As a result, for a dislocation loop with ${\bf b}={\bf e}_x$, according to the outlined rules, we find modes bound {\it only} to the the edge-dislocation parts. As both these planes contribute the midgap states, we expect a double Kramers pair of the propagating metallic states, which can be verified by numerical computations \cite{kbt2014}. We emphasize, however, these modes can mix in the dislocation loop, since no symmetry relates the momenta $M$ and $R$ giving rise to them. Hence, this directly amounts to consistency with the above notions based on the weak indices. Nevertheless, in terms of principle we thus observe that the richer ${\bf K}\text{-}{\bf b}\text{-}{\bf t}$ rule consistently unveils the formation mechanism of the dislocation modes, whereas the the underlying symmetry directly conveys the stability. 

Finally, a similar compatibilty is retrieved in the remaining scenarios in which the phase can be viewed as a layered phase. Specifically, considering the tetragonal $p4_{X',R}$ ($p4_{Y',R}$) phase by deforming the cubic lattice in the ${\bf e}_y$ (${\bf e}_x$) direction with the corresponding band-inversions at $(\pi,0,\pi)$ $[(0,\pi,\pi)]$ and  $(\pi,\alpha\pi,\pi)$ $[(\alpha\pi,\pi,\pi)]$ momenta (Fig. \ref{figure::cvone}(D)), it can be verified that no dislocation modes appear in the $p4_{X',R}$ phase for the same dislocation configuration. In contrast, in the $p4_{Y',R}$ phase, only the band-inversion at momentum $(\alpha\pi,\pi,\pi)$ contributes a $\pi$ flux, thus yielding  modes for both types of dislocations.
This once more reconfirms that the above view naturally includes the previously obtained results, although the generalization itself has a direct impact in terms of the outlined topological indexing of the underlying phase and characterization of the associated dislocation modes. 

\section{Self-organized semimetals on extended defects\label{sec::GBmetals}}
The described dislocation modes are physically interesting due to their unusual spin charge separated quantum numbers. The halve number of degrees of freedom is reminiscent of the usual edge states. Nonetheless, as they parameterize one spatial dimension less, new effects can be attained. As a particular example it can be shown that such modes may hybridize into semimetallic band structures across extended defects, such as grain boundaries \cite{Slager2016}. Most importantly, these grain boundary metals can be linked to exotic transport properties, which in some instances can directly be traced back to field theoretical anomalies that have an intricate relation to edge states of the parent TBI.

\subsection{Formation mechanism}
\begin{figure}
\includegraphics[width=0.75\columnwidth]{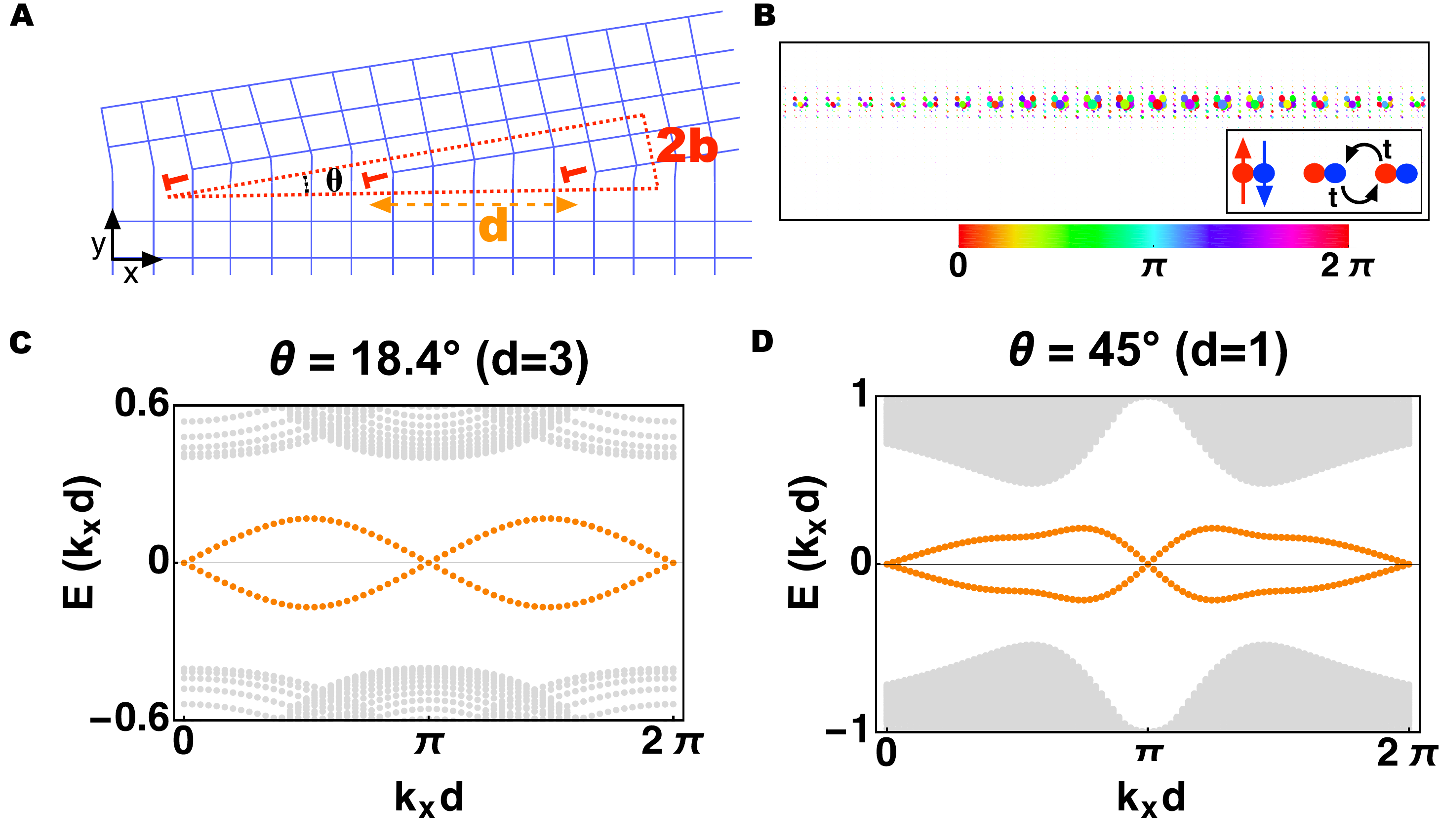}
\caption{\label{Fig::c6TW1}
Grain boundary semimetal on a 1D grain boundary in a translationally active 2D TBI ($M$-phase of the BHZ model \cite{Bernevig2006}), adopted from \cite{Slager2016}. ({\bf A}) Schematic illustration of a GB. The coordination discrepancy due to the angular mismatch $\theta$ of lattice basis vectors results in an effective array of dislocations. These dislocations, having spacing $d=|\mathbf{b}|/\tan \theta$, are described by Burgers vector $\mathbf{b}$ and marked by a 'T' symbol. ({\bf B}) Real space numerical tight-binding calculation of the low-energy states in the presence of a GB. The spinon degrees of freedom bound to the dislocations hybridize with tunneling amplitude $t$, which gives rise to an extended 1D state along the GB. The radii of the circles indicate the amplitude of a wave function, associated with the node at $k_x=\pi/d$ of panel (C), while the colors indicate the phase. ({\bf C}) The characteristic mid-gap bow-tie dispersion of the spinon bands (orange) and bulk bands (grey), corresponding to GB opening angle $\theta=18.4\degree$ for which isolated dislocations along the GB are well defined, as the function of the momentum $0 \leq k_x \leq 2\pi/d$ along the GB. (D) The same plot for the maximal opening angle $\theta=45\degree$ and a simulation that only involves a coordinate mismatch. While the picture of isolated dislocations breaks down and the simple nearest-neighbor hybridization based sinusoidal dispersion is lost, the defining TRS semi-metal dispersion with the nodes at the time-reversal symmetric momenta survives.}
\end{figure}

Consider in this regard a 2D translationally active phase so that individual dislocations bind a single Kramers pair of modes in the presence of a grain boundary (GB). 
Although it has been long known that interfaces connecting TBIs can accommodate modes in the presence of a protecting mirror symmetry \cite{Takahashi11},  grain boundaries have a real topological status in the crystal geometry \cite{kleinert, Gb}. They arise at the interface of two crystal regions whose lattice vectors are misaligned by an angle $\theta$, as illustrated in Fig. \ref{Fig::c6TW1}(A), and for small opening angles may be viewed as a stack of lattice dislocations described by a Burgers vector $\mathbf{b}$ arranged on an array of spacing $d=|\mathbf{b}|/(\tan \theta)$. As result,  GBs provide for an experimentally viable setting for the mobilization of the previously derived dislocation modes,  serving as building blocks for new states of matter without requiring lattice symmetries or fine-tuning. Indeed, using numerical means it is readily shown that a grain boundary supports a 1D propagating localized mode. Specifically, this state appears in the energy spectrum as a gapless mid-gap Õbow-tieÕ band that exhibits two non-degenerate nodes, reminiscent of the valleys in graphene. These nodes appear at the TRS momenta along the GB and persist for all GB angles including the maximal opening angle of $\theta = 45\degree$. Hence, for a neutral GB where the chemical potential is at the nodes, a one-dimensional semi-metal emerges.

The microscopic origin of these grain boundary metals lies in the hybridization mechanism of the dislocation modes. 
When brought into proximity, these effective spinon modes hybridize between dislocations giving rise to a nearest-neighbor hopping term of magnitude $t$. Correspondingly, a translationally invariant GB can then be modeled in terms of a simple TRS tight-binding model by virtue of the incompressibility of the parent system. This results in a momentum space Hamiltonian of the general form
\begin{equation}\label{eq::c6main}
H(q)=\mathbf{h}(q) \cdot \bm{\sigma}.
\end{equation}
The specific form can subsequently be determined from calculating all hybridization terms and projecting to the spinon subspace. More importantly, as the Pauli matrices $\bm{\sigma}$ act in {\it spin space}, the presence of time reversal symmetry in turn requires $H(-q)=-H(q)$, resulting in the vanishing of $H(q)$ at the TRS momenta $q=0$ and $q=\pi$. In the full TBI where the presence of a GB oriented in $\hat{x}$-direction has translational symmetry only with respect to the dislocation spacing $d$, these TRS momenta correspond to $k_x=0$ and $k_x=\pi/d$. Hence, the spinons hybridize into two helical bands, one for each spin component, that are odd in momentum and feature TRS protected nodes at the two TRS invariant momenta, fulfilling the fermion doubling theorem \cite{Nielsen1981a, Nielsen1981b}  by virtue of the nodal separation in momentum space. 

This can be further motivated by the previously mentioned Volterra approximation. Recall that in this case the dislocation modes are seen as solitons in the necessarily present edge state spectrum on each side of a hypothetical cut as the hoppings acquire a phase shift across the part comprising the Volterra plane \cite{Ran2009}. The generality with respect to the underlying topology of the parent state is then recovered by the $\mathbf{K}$-$\mathbf{b}$-$\mathbf{t}$ rule. Namely, the interplay of $\mathbf{t}$ and $\mathbf{b}$ unambiguously coveys the edge states, having velocity $v$, relevant to construct the spinon modes as function of the particular plane and band inversion momentum $\mathbf{K}_{inv}$. Under the Volterra approximation the projection operators $P$, however, take  such a simple form that the explicit form of Eq. \eqref{eq::c6main} can be determined analytically. Specifically, in a regular grain boundary, being a stack of dislocations, the adjacent edge states couple via a translationally and TRS invariant Hamiltonian $H_{GB}=2t[\cos(q) \mu_{x}\otimes\sigma_{0} + \sin(q) \mu_{y}\otimes\sigma_{0}]$ \cite{Slager2016}. Here $q$ spans the reduced Brillouin zone $ k_x=0 \geq q \geq  k_x=2\pi/d$, the $\mu$ matrices parametrize the two pairs of edge states on each side of the hypothetical cut [see Section \ref{sec::weak}] and we only took into account nearest neighbor hopping. Projecting this coupling into the spinon subspace, one finds that the spinons acquire a dynamics described by $P (H_0 + H_{GB}) P = 2t\sin(q) \tilde{\sigma}_{z}$, where $\tilde{\sigma}_{z}$ entails an effective spin operator in the spinon subspace. Moreover, additional Rashba terms, such as $H_R=\alpha k_y \mu_{z}\otimes\sigma_{y}$, inversion breaking terms $H_I=\sum_i m_i \sigma^i \mu^y$ (here $m_i$ can be either constant or proportional to $\cos k_x$) and TRS breaking Zeeman terms $H_B=\sum_i b_i \mu_{0}\otimes\sigma_{i}$ can naturally be included in this analysis. Specifically, projecting these into the spinon subspace, yields an minimal effective model of the form 
\begin{equation}\label{Eq::AScience2Deff}
	H_{2D} = \frac{m_y \alpha + m_z v}{\sqrt{v^2+\alpha^2}} + \left( \frac{vb_z + \alpha b_y}{\sqrt{v^2+\alpha^2}} + t\sin k_x \right) \tilde{\sigma}^y,
\end{equation}
which can be directly corroborated by numerics \cite{Slager2016}. 

The above expression \eqref{Eq::AScience2Deff} shows that the key property determining the response to TRS and inversion breaking is the spin texture. A Rashba coupling that is not strong enough to close the bulk band gap preserves the semi-metallic nodes. Moreover, inversion breaking terms in general shift the nodes to different energies, but, in contrast to interfaces of TBIs with different velocities \cite{Takahashi11}, cannot gap them out. Hence, we emphasize that the formed semi-metal has the usually desired property that the Fermi level is fixed at the nodes {\it for free} when the parent TBI has inversion symmetry. Moreover, TRS breaking terms gap the spectrum, but only if they do not anti-commute with the projection operators of the dislocation spinons and only once their magnitude is comparable to the bandwidth resulting from the hybridization. Furthermore, it can also numerically be established that no moderate random disorder can gap the system since the topological edge states underlying the hybridizing dislocation modes persist as long as the disorder does not drive the bulk out of the translationally active phase. The self-organized state on the GB thus finds its topological stability ultimately in stability of the parent state. This also implies that perfect translational invariance along the GB is not a necessary condition for the stability of the nodal band structure. Finally, similar
to chemical potential disorder that deforms dislocation mode
wave functions, random bends along the GB result in random
tunneling couplings $t$. In terms of the low-energy theory around the nodes, this gives rise to a random gauge field that only randomly shifts the cones \cite{Foster06}. As long as this shift is smaller than the separation $\pi/d$ of the cones in the BZ, the semi-metallic behavior is stable.

\begin{figure}
\includegraphics[width=0.75\columnwidth]{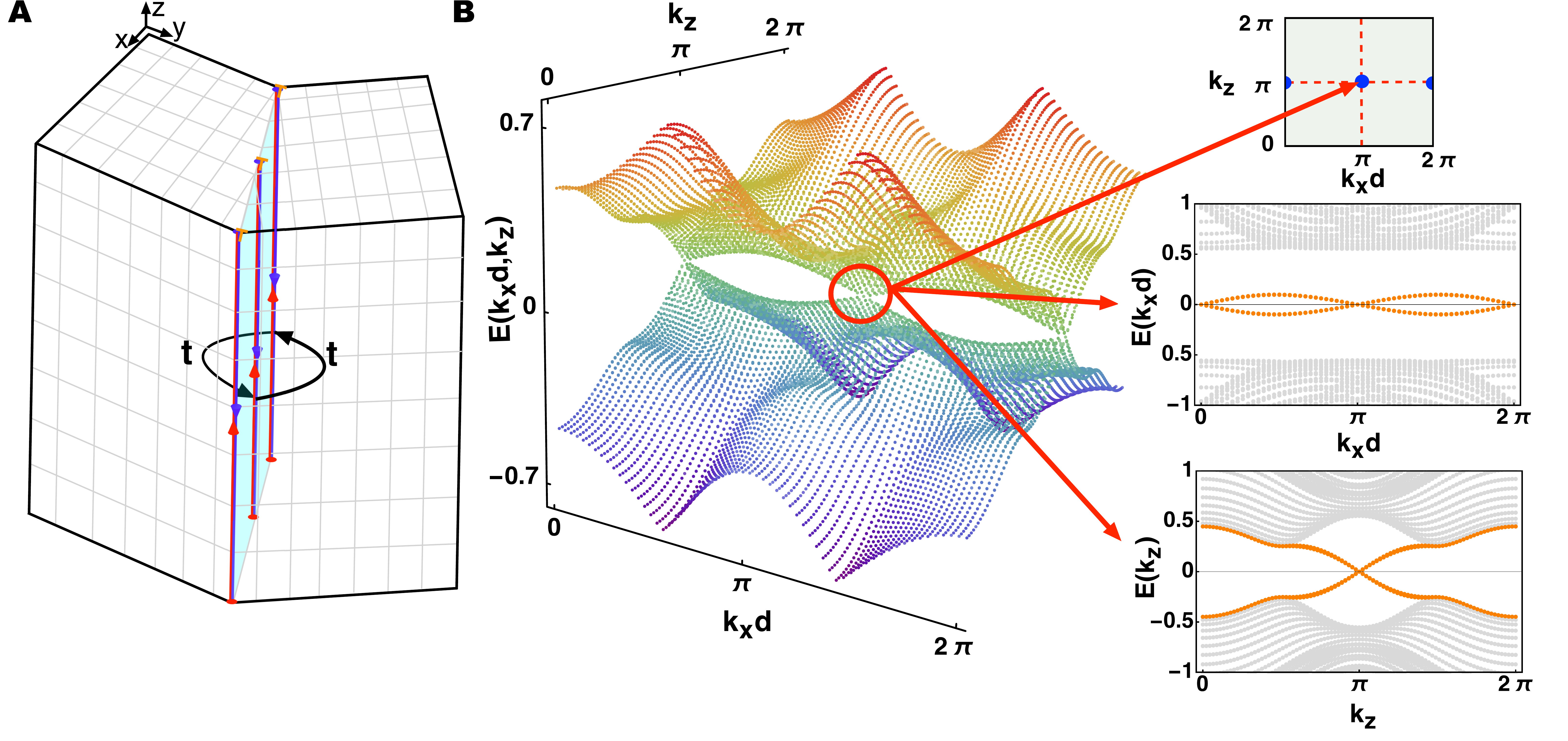}
\caption{\label{Fig::two}
Grain boundary metal localized on a 2D GB in a translationally active 3D topological band insulator ( $R$-phase on a cubic lattice), adopted from \cite{Slager2016}. ({\bf A}) Schematic illustration of the 2D GB, which is realized from a sheet of hybridizing parallel 1D edge dislocation lines. Each edge dislocation hosts a pair of propagating helical modes localized along the dislocation cores that cross at $k_z=\pi$. ({\bf B}) On a GB of opening angle $18.4\degree$, the hybridization of the helical modes results in the semi-metal band structure with two anisotropic pseudo-relativistic fermions at $(k_x,k_z)=(0,\pi)$ and $(\pi/d,\pi)$. Along the GB we recover the same mid-gap bow-tie dispersion as in the 2D case (shown top right for $k_z=\pi$), while along the edge dislocations the hybridized modes (orange) still flow into the bulk bands (gray) as a function of $k_z$ (bottom right for $k_x=\pi/d$). }\end{figure}

More importantly, the outlined mechanism directly applies in 3D translationally active TBIs, where a GB consists of a 2D sheet of parallel 1D dislocation lines, as illustrated in Figure \eqref{Fig::two}A. In a translationally active phase each dislocation binds a Kramers pair of helical modes \cite{Ran2009,Tteo2010, kbt2014} and a similar reasoning thus reveals that the hybridization process leads to the anticipated grain boundary metal. This semimetal is very reminiscent of graphene, having two linearly dispersing cones at the two TRS momenta $(k_x,k_z)=(0,\pi)$ and $(\pi/d,\pi)$, although the velocities $v$ along and $t$ perpendicular to the dislocation lines, respectively, are now generally anisotropic. Concretely, Eq. \eqref{Eq::AScience2Deff} may be readily generalized to 
\begin{equation}
	H_{3D} = m_y + v k_z \tilde{\sigma}^z +  \left(  h_y + t\sin k_x \right) \tilde{\sigma}^y.
	\label{Eq::Hameff_3d}
\end{equation}
We note that the resultant anisotropy for the cones in both directions is reduced for larger opening angles $\theta$ due to an enhanced overlap between the helical modes whose hybridization underlies the self-organized GB state. These cones are separated in momentum space and are degenerate in energy if the inversion symmetry with respect to the GB plane is preserved ($m_y=0$). Finally, it is evident that the stability arguments pertaining to the 2D case similarly translate to the 3D scenario. 

\subsection{Experimental consequences}
\begin{figure}
\includegraphics[width=0.6\columnwidth]{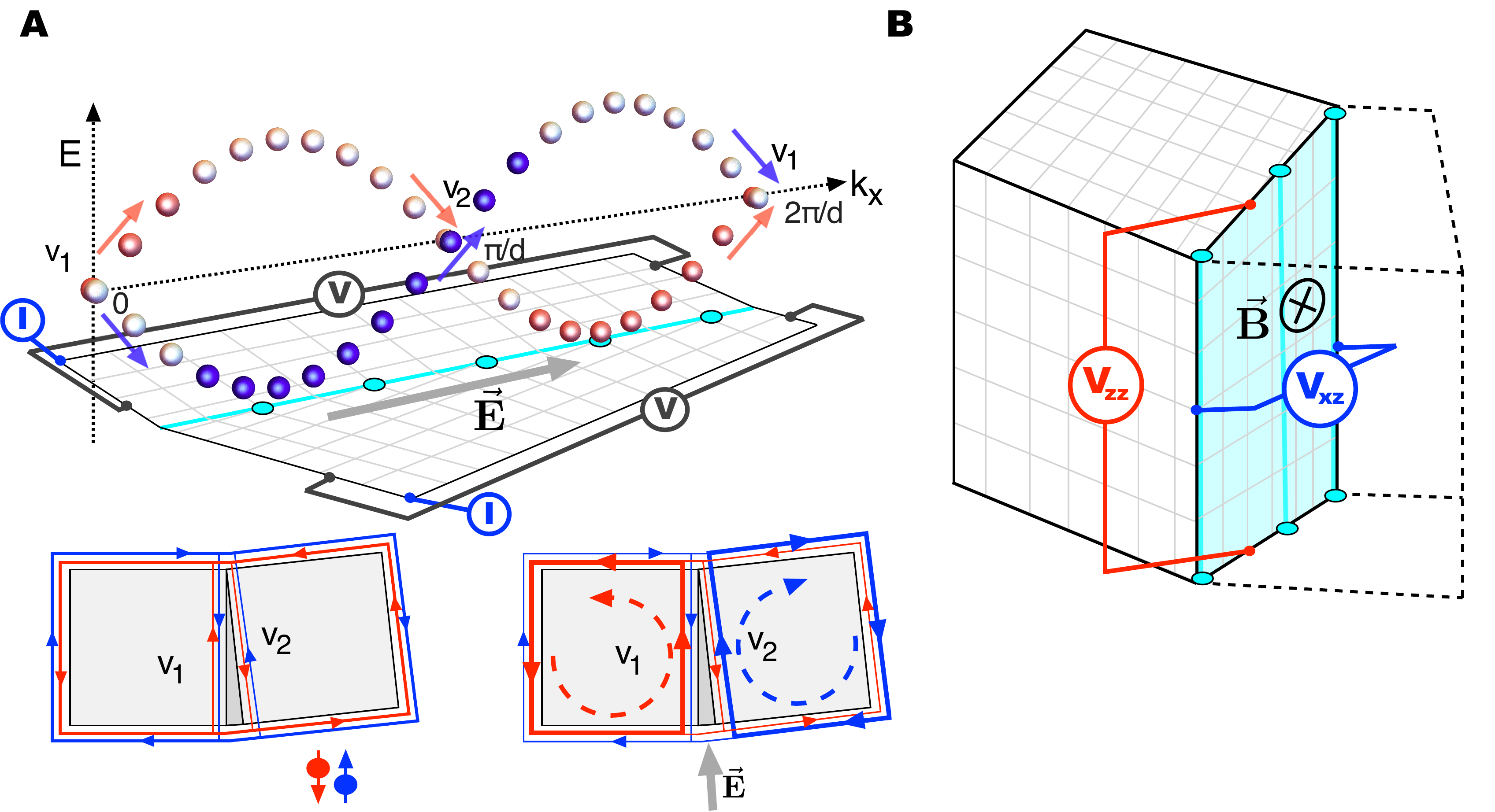}
\caption{\label{fig:exp} Proposed experimental setups to identify the signatures of the grain boundary semimetals (adopted from Ref. \cite{Slager2016}). ({\bf A}) One-dimensional bow-tie dispersion of the spinon semi-metal on a 1D GB implies a parity anomaly when an electric field $\vec{E}$ is applied along the grain boundary. The arrows indicate the shift in the spectrum from one valley to the other: Valley $V1$ accumulates an excess of spin down (red), while valley $V2$ accumulates an excess of spin up (blue). These valleys can be associated with two helical two-mode channels that connect the helical edge states from opposing surfaces. When $\vec{E}$ is applied along the GB, a current for both spin orientations is driven parallel to it. At the GB edges this current flows into the edge states that propagate to opposite directions resulting in a doubled spin Hall effect -like spin imbalance of the edge states on the two sides of the GB. Measuring this imbalance is the hallmark signature of the valley anomaly exhibited by the spinon semi-metal.  ({\bf B}) The spinon semi-metal on a 2D GB features an odd integer Hall effect with transverse conductivity $\sigma_{xz}=(2n+1)e^{2}/h$ in the presence of the perpendicular magnetic field $\vec{B}$. In the absence of external fields, another signature is provided by the diagonal ballistic optical conductivity of $\sigma_{zz}=(\pi/4) e^2/h$, which is half that of graphene. }
\end{figure}

Although the formation mechanism of the above mentioned spinon semimetals poses an intriguing process, the consequential relevance is ultimately fueled by possible experimental signatures. 
To detect the semi-metal on the 1D GB inside a 2D TBI, one could carry out a two terminal transport measurement analogous to the one used to detect edge states in a 2D quantum spin Hall insulator \cite{Koning2007}. When an electric field ${\bf E}$ is applied along the GB, the two cones should result in a conductance of $\sigma=4e^2/h$, i.e. twice the value measured for the QSH edge states.
A more dramatic result of the helical bow-tie dispersion along a 1D GB is the direct correspondence with a parity anomaly \cite{Nielsen-Ninomiya1983} for each spin projection. This, in the presence of two chiral cones, gives rise to a {\it valley anomaly}, due to an intricate relation with the edge states of the parent TBI. As illustrated in Fig.\ \ref{fig:exp}(A), an electric field ${\bf E}$ along the GB generates a current for both spin components that results in one valley having excess of spin up and the other spin down. As the valleys can be associated with two distinct channels through which the helical edge states can flow from one GB termination surface to the other, the valley anomaly gives rise to the spin polarization of the edge currents at different sides of the GB. This spin imbalance as a result of applied electric field is proportional to the hybridization strength $2t$ and represents a hallmark signature of the valley anomaly that originates from the parity anomaly for each spin component of the GB semi-metal. To detect it, one can carry out a two-terminal edge transport measurement illustrated in Fig.\ \ref{fig:exp}(A). When the two edges connected by the GB are biased by voltage $V$, one expects net current $I \sim 2t \frac{4e^2}{h} V$ reflecting unequal spin currents. Alternatively, the valley anomaly may be detected by measuring the magnetic moment due to spin imbalance \cite{Nowack2013} or using Kerr rotation microscopy \cite{Kato2004, Sih2005}. 

Although the "isospinless graphene" on a 2D GB can also be viewed as two helical channels that connect the surface states of the 3D TBI and may give rise to a (non-quantized) spin-pumping effect, it is experimentally most conveniently detected via distinct "half-graphene" -like signatures. When a magnetic field is applied perpendicular to the GB in the setup shown in Fig.\ \ref{fig:exp}(B), the non-degeneracy of each cone implies a contribution of $(n+1/2) e^2$/h to the Hall conductivity, which in turn implies odd-integer quantum Hall effect with total Hall conductivity  $\sigma_{xz}=(2n+1) e^2$/h. In addition, in the absence of the magnetic field, the measured ballistic optical conductivity is $\sigma_{zz}=(\pi/4) e^2/h$, which is half the value measured in graphene \cite{ballistic-minimal-conductivity}. Finally, the semimetal can also be expected to exhibit edge states that should contribute additional Fermi arc-like features to the surface states of the parent 3D TBI \cite{Grushin15}.

\section{Road to reality?\label{sec::Experimental}}
Although the primary interest around TBIs has been revolving around the edge state signatures, as also reflected by the enormous interest in the second generation bismuth/antimony alloyed TBIs exhibiting a simple single cone around the $\Gamma$ point \cite{Xia2009, Chen2009, Zhang2009}, the very first discovered 3D TBI {\it is} actually of the transitionally active kind \cite{Hsieh2007}. Moreover, apart from this  $\text{Bi}_{x}\text{Sb}_{1-x}$ compound, 
the list of translationally active phases is in fact growing both in terms of prediction and experimental verification. Indeed, the fully isotropic $R$-phase, for example, is believed to be realizable in electron-doped BaBiO$_2$ \cite{BinghaiYan2013}. Similarly, the predicted topological Kondo insulating material SmB$_6$ has band inversions at the three $X$ points \cite{Xu2014, SmB6} and the recently discovered bismuth iodide compounds  feature a transitionally active phase with band inversions at the $Y$ points \cite{Autes2015}. In this regard, topological crystalline compounds, such as Sn-based materials  \cite{fuNatComm2012, Dziawa2012, Ando2012}, are also of direct interest for realizing the outlined physical mechanisms as long as the protecting symmetry is respected. 
 
\begin{figure}
\includegraphics[width=0.3\columnwidth]{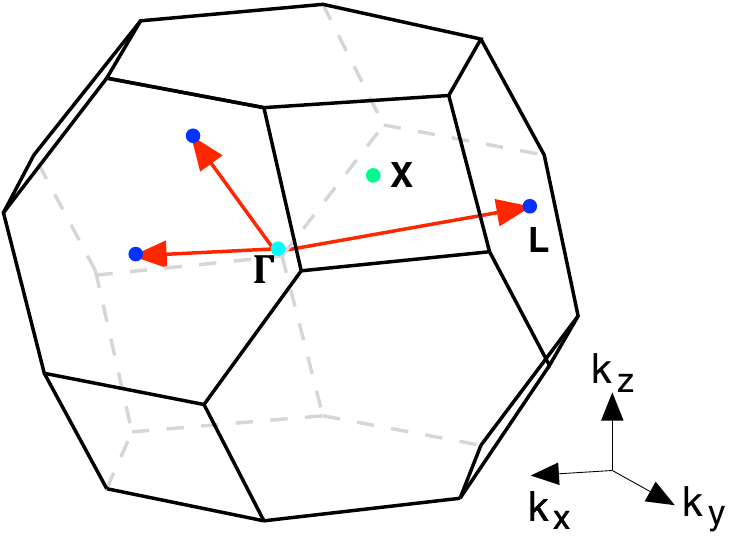}
\caption{\label{fig::BiSb} Brillouin zone of $\text{Bi}_{x}\text{Sb}_{1-x}$. The reciprocal vectors connect the turquoise $\Gamma$ and blue $L$ points. As the points of band inversion entail the $L$ points, every dislocation with a primitive Burgers vector results in dislocation modes. In contrast, if the dislocation's Burgers vector amounts to a combination of two reciprocal vectors no modes are formed according to the outlined criteria.   }
\end{figure}

However, apart from the encouraging activity on the material side, 
the first experimental signatures of dislocation mode physics in $\text{Bi}_{x}\text{Sb}_{1-x}$,  as reported in Ref. \cite{Hamasaki2017}, may be perceived as even more inspiring. To be more concrete, in the above nomenclature, the topological phase of  $\text{Bi}_{x}\text{Sb}_{1-x}$ entails a $T-r\bar{3}_L$ phase \cite{Slager2016}. This material has space group $R\bar{3}m$ and hence, by virtue of the inversion symmetry, the sign structure of the $\delta_{i}$ configuration, c.f. Eq. \eqref{eq::nu}, is simply set by the parity. Of the TRI momenta, comprising the $\Gamma$ point, three inequivalent $X$ points and three $L$ points, the $\delta_i$'s at the $L$ points are then of opposite sign. In fact the reciprocal vectors are those that connect the $\Gamma$ and $L$ points (Figure \eqref{fig::BiSb}). As a result, for all combinations of planes, characterized by ${\bf t}$, and Burgers vectors ${\bf b}$ modes are anticipated. Indeed, any dislocation with a primitive Burgers vector "activates" a single point, resulting in a single pair of modes. Conduction measurements in samples with such dislocations indicate that these extra helical channels are indeed formed, contrasting the case when the formation criteria are not met [i.e having dislocations with a Burgers vector of the form $(1,-1,0)$] significantly \cite{Hamasaki2017}. With regard of the formation process as outlined by the ${{\bf K}\text{-}{\bf b}}\text{-}{\bf  t}$ rule, we note that comparing this results with dislocations that have a Burgers vector of the form $(1,1,1)$ would also be interesting, as in the latter case one would expect more modes contributing to the conduction. 

Although these first indications should firstly be further explored, as a next step it should then be conceivable to grow a relative regular grain boundary by joining two ribbons \cite{Lima2016}. Especially the fabrication of a 2D
GB in a 3D TBI could be facilitated by previous experimental knowledge on bi-crystal techniques \cite{Ikuhara2011}. Additionally, standard lithography can then be used to
fabricate Hall bars for the according Hall measurements. Given the outlined distinct experimental features, this might then culminate in the ultimate experimental verification of the above as well as new effects.

 \section{Concluding remarks\label{sec::Conclusions}}
We have provided a perspective on some of the exciting features of the bulk of topological band insulators. Although TBIs are usually characterized by the presence of edge states, the topological entity of the bulk can directly  be characterized by the criterion whether a charge-0 time-reversal symmetric $\pi$ flux mode binds
a Kramers doublet of modes. In spite of being arguably not very viable from a experimental perspective, the $\pi$ flux modes are nonetheless intricately related to the action of omnipresent dislocations under specified conditions. These conditions directly correspond to the role of dislocations as being the unique probes of the translational symmetry breaking by the necessarily present underlying crystal lattice. As such, these criteria instigate an indexing that takes these space group symmetries into account and whose resulting classes are directly probed by the appearance of dislocation modes. 

Moreover, dislocation modes also harbor interesting physics in themselves. Indeed, they share many of the interesting features of the edge states, but, by virtue of living in one spatial dimension less, these properties can be utilized in new schemes. In particular, dislocations modes can hybridize into self-organized semimetals that exhibit distinct transport phenomena and an intricate relation to the edge states of the parent TBI. In the presence of inversion symmetry these semimetals even feature a spectrum with nodes at the Fermi energy for free. 

Given the timely experimental progress on both the material and experimental side, the above anticipated physics could well become accessible in a real setting. Moreover, new consequences, utilizing the spin-charge separated nature and the dimensionality of the modes, should results in new proposals. In the latter case one could specifically also think of braiding dislocation modes.
At the very least, all these results pose a new agenda beyond the very successful, but by now rather explored, efforts involving the usual edge states.

{\it -Acknowledgements} The above perspective has been subject of Ph.D thesis work \cite{slager2016symmetry}, which was supported by the Dutch Foundation for Fundamental Research on Matter (FOM). The author thanks V. Juricic and J. Zaanen for reading (parts of) the manuscript and providing comments. The author readily acknowledges all collaborators of the various projects underlying this perspective.

\bibliographystyle{apsrev4-1}
\bibliography{references}

\end{document}